\documentclass[preprint]{aastex}

\newcommand{\thetaph}{\theta_{\rm ph}}
\newcommand{\phiph}{\varphi_{\rm ph}}

\slugcomment{Draft: \protect\today}

\shorttitle{Accretion Disk Spectra of Ultra-luminous Compact X-ray Sources}
\shortauthors{Ebisawa et al.}

\begin{document}


\title{Accretion Disk Spectra of the
Ultra-luminous X-ray Sources in Nearby Spiral
Galaxies and Galactic Superluminal Jet Sources}


\author{Ken Ebisawa\altaffilmark{1}}
\affil{code 662,  NASA/GSFC, Greenbelt, MD 20771, USA \\
and \\INTEGRAL Science Data Centre, chemin d'\'Ecogia 16, Versoix, 1290 Switzerland
}
\email{ebisawa@obs.unige.ch}
\author{Piotr \.Zycki}
\affil{N. Copernicus Astronomical Center, Bartycka 18, 00-716 Warsaw, Poland}
\author{Aya Kubota}
\affil{Institute of Space and Astronautical Science, 3-1-1 Yoshinodai,
Sagamihara, Kanagawa, 229--8510 Japan}
\author{Tsunefumi  Mizuno}
\affil{Stanford Linear Accelerator Center, 
 2575 Sand Hill Road, M/S 43A,
 Menlo Park, CA 94025, USA}
\and
\author{Ken-ya Watarai}
\affil{Yukawa Institute for Theoretical Physics,  Kyoto University,
Sakyo-ku, Kyoto, 606-8502 Japan}

\altaffiltext{1}{Universities Space Research Association.}


\begin{abstract}
Ultra-luminous Compact X-ray Sources (ULXs)
in nearby spiral galaxies and  Galactic superluminal
jet sources share the common spectral characteristic that they have
unusually  high  disk temperatures which cannot be explained
in the framework of the standard optically thick accretion disk 
in the Schwarzschild metric.  
On the other hand, the standard  accretion disk around the Kerr black hole might explain the 
observed high disk temperature, 
as the inner  radius of the Kerr disk gets smaller and
the disk temperature can be  consequently higher.
However, we point out  that the observable Kerr disk spectra  becomes significantly harder than
Schwarzschild disk spectra 
{\em only when}\/ the disk is highly inclined.
This is because the emission from the innermost part of the accretion disk
is Doppler-boosted for an edge-on Kerr disk, while hardly  seen for a face-on disk.
The Galactic superluminal jet sources are known to be highly inclined systems, thus their energy 
spectra may be  explained with the standard Kerr disk with 
known black hole masses.
For ULXs, on the other hand, the standard   Kerr disk model seems implausible,
since it is highly unlikely that their accretion disks are preferentially inclined,
and, if edge-on Kerr disk model is applied, the black hole mass becomes
unreasonably large ($\gtrsim 300 M_\odot$).
Instead, the slim disk (advection dominated optically thick disk) model
is likely to explain the observed super-Eddington luminosities,  hard energy spectra,
and spectral variations of ULXs.
We suggest that ULXs are accreting  black holes with a few tens of solar mass, 
which is not unexpected  from the standard  stellar evolution scenario, and that
their X-ray emission is from the slim disk shining at super-Eddington luminosities.

\end{abstract}




\section{Introduction}

Ultra-luminous compact X-ray sources (ULXs) have been found in nearby spiral Galaxies,
with typical 0.5 -- 10 keV luminosities  $10^{39}$ to $10^{40}$ erg s$^{-1}$
(e.g.,  Fabbiano 1988; 
Colbert and Mushotzky 1999; Makishima et al.\ 2000; 
Colbert and Ptak 2002; Foschini et al.\ 2002).
These luminosities are too small for AGNs, and most ULXs are in fact  located
significantly far from the  photometric center of the galaxies.  
On the other hand,  ULXs are too luminous to be considered as the same class of 
the compact binary X-ray sources in our Galaxy whose luminosities are
almost always $\lesssim 10^{39}$ erg s$^{-1}$.

Significant time variations have been detected from ULXs, and 
their energy spectra are successfully modeled with emission from 
optically thick accretion disks
 (Okada et al.\ 1998; Mizuno et al.\ 1999; Kotoku et al.\ 2000;
Makishima et al.\ 2000; Mizuno, Kubota and Makishima 2001),  as is the case for 
the ``High'' state (= Soft-state)  of Galactic
black hole candidates  (e.g., Tanaka and Lewin 1995).
In addition,  discovery of
bimodal-type  spectral transitions  
(Kubota et al.\ 2001) and orbital modulations (Bauer et al.\ 2001; Sugiho et al.\ 2001)
from several ULXs 
further demonstrate their  resemblance with  Galactic black hole candidates.
These observational facts  suggest that ULXs are moderately massive
black holes,  which may be scale-up versions of the Galactic black holes.
So that the observed luminosities  of
 ULXs,  $\sim 10^{40} $ erg s$^{-1}$,   
do not exceed the Eddington limit 
$L_{Edd} = 1.5 \times 10^{38} \,(M/M_\odot)$ erg s$^{-1}$,
the black hole mass   has  to be as large as or greater  than    $\sim$100 $M_\odot$,
assuming  isotropic emission.
How to create such ``intermediate'' mass black holes, if truly exist, 
is an intriguing question (e.g.,  Ebisuzaki et al.\ 2001).
On the other hand, if X-ray emission is unisotropic,
the black hole mass is not required to be so large.
For example, a beaming model for ULXs has been proposed
(King et al. 2001; K\"ording, Falcke and Markoff 2002); however,
presence of a bright nebula surrounding  M81 X-9 
(Wang 2002) suggests that the nebula is  powered by the central X-ray source and that
strong X-ray beaming is rather unlikely.
Another possibility to solve the super-Eddington problem through  unisotropic emission  
is a geometrically
thick accretion disk (Watarai, Mizuno and Mineshige  2001), which we favorably consider later
 (section \ref{slim-disk}).

GRS 1915+105 and GRO J1655--40 are the two well-known  Galactic 
superluminal jet sources and  established
black hole binaries with  reliable mass measurements 
($M 6.3 \pm 0.5 M_\odot $ for GRO J1655--40 [Green,  Bailyn and Orosz 2003] and 
$\approx14 M_\odot$ for GRS 1915+40 [Greiner,  Cuby and  McCaughrean 2001]).
If  soft X-ray energy spectra of these sources are fitted
with an optically thick accretion disk model,
their characteristic disk temperatures are 
 $\sim $1.3 -- 2.0 keV (e.g., Belloni et al.\ 1997;
Zhang et al.\ 1997; Zhang, Cui and Chen 1997; Tomsick et al.\ 1999).  These values  are 
systematically higher than 
those of ordinary and  well-studied soft-state black hole candidates such as Cyg X-1 and   
LMC X-3,
whose disk temperatures are  almost always less than $\sim$ 1 keV (e.g., Tanaka and Lewin 1995).

Okada et al.\ (1998) found that a 
luminous ULX in IC342
($\sim 2 \times 10^{40}$ erg s$^{-1}$; ``Source 1'') 
has  unusually high disk temperature  ($ \sim$1.7 keV), which is 
similar to the  accretion disk spectra of Galactic superluminal jet sources,
and   pointed out that such a  high disk temperature 
cannot be explained in the framework of  the ``standard'' accretion disk
around a Schwarzschild  black hole  (see section \ref{problem} and \ref{ApplicationIC342}).
The ``standard'' disk denotes the situation 
that all the  gravitational energy release is converted to thermal radiation,
in contrast to the ``slim'' disk 
 which is an optically and geometrically
thick disk with  dominant energy
advection (section \ref{slim-disk}).
Makishima et al.\ (2000) summarized  ASCA observations  of seven ULXs, and
concluded that the unusually high accretion disk temperature 
is a common spectral property of  ULXs.
King and Puchnarewicz (2003) pointed out that
ULXs, as well as ultrasoft AGNs, violate the
apparent blackbody-temperature and luminosity
relationship which is required not to exceed the Eddington limit.
Zhang, Cui and Chen (1997) and Makishima et al.\ (2000)  suggested 
that the unusually high disk temperature 
of the Galactic superluminal jet sources and ULXs may 
be explained if they harbor fast rotating  black holes,
since inner edge of the accretion disk gets closer to the black hole 
in the Kerr geometry and the disk can be hotter.
On the other hand, Watarai et al.\ (2000) and Watarai, Mizuno and Mineshige (2001)
proposed that, instead of the standard disk,
the slim disk model may explain the X-ray energy spectra of Galactic
superluminal jet sources and ULXs.

In this paper, we focus on the ``too-hot disk'' problem of ULXs and Galactic superluminal
jet sources. We apply standard  accretion disk models in Newtonian, Schwarzschild, and Kerr
cases, and discuss how the relativistic effects and the disk inclination 
affect the black hole mass and the mass accretion rates obtained
from the model fitting.
We see that observed hard spectra of Galactic superluminal jet sources
may be explained by strong relativistic effects in the 
 highly inclined  standard Kerr accretion disk.
On the other hand, the hard spectra and the super-Eddington problem of ULXs 
are difficult to  explain in the framework of the standard accretion disk model.
We see that the slim disk model may 
explain the observed super-Eddington luminosities and hard energy spectra of ULXs
more naturally.

\section{Characteristics of the Standard Accretion Disk Model}
Before quantitative discussion of the observed energy spectra and 
accretion disk parameters, we summarize important 
observational characteristics
of the standard accretion disk model.

\subsection{Disk Temperature}
\label{problem}
Let's assume a standard optically thick accretion disk 
(Shakura and Sunyaev 1973)
in 
the Schwarzschild metric,
in which case the last stable orbit around the non-rotating black hole
is  $ 6\; r_{g}$, where the gravitational
radius $r_{g} \equiv GM/c^2$ is defined.
We identify the last stable orbit with the  innermost disk radius, $r_{in}$.
The radial dependence
of the ``effective'' temperature of an  optically thick accretion disk
may be written as,
\begin{equation}
 T_{eff}(r)= \left\{\frac{3GM\dot M}{8\pi\sigma r^3} R_R(r/r_{in})\right\}^{1/4}, 
\label{Teff}
\end{equation}
where $R_R$ is $\left(1-\sqrt{r_{in}/r}\right)$ in the Newtonian case,
otherwise  includes  additional  general relativistic correction
(see e.g., Krolik 1999).
The effective temperature is zero at $r=r_{in}$,
peaks at $r \approx 8 \: r_g$, and decreases with $\propto r^{-3/4}$ outward.

Because of the Comptonization in the disk atmosphere, the color temperature of the
local emission becomes higher than the effective temperature,
and their ratio is almost constant at $\approx 1.7$ (section \ref{local}).
Therefore, the observable disk color temperature  also peaks at $r \approx 8 r_g$,
and takes the maximum values,
\begin{equation}\label{Tmax}
T_{col}^{(max)}  \approx (1.0 - 1.3) \; {\rm keV} 
\left(\frac{T_{col}/T_{eff}}{1.7}\right)\!\!
\left(\frac{\dot M}{\dot M_C}\right)^{1/4}\!\!\!\!
\left(\frac{M}{7 \,M_\odot}\right)^{-1/4},
\label{Tcolmax}
\end{equation}
where we define the critical mass accretion rate $\dot M_{C} \equiv
2.9 \times 10^{18} (M/M_\odot)$ g s$^{-1}$,
so that $\dot M = \dot M_{C} $  gives the Eddington luminosity
(see appendix \ref{efficiency_appendix}).\label{M_C_definition}
The small temperature variation reflects the  inclination effects
(from face-on to $i=80^\circ$), such
that inclined Schwarzschild disks have slightly harder spectra (section 
\ref{RelativisticEffects} and Appendix \ref{GRADbug}).

From equation (\ref{Tcolmax}), we can see that an optically thick accretion
disk around a 7 $M_\odot$  or 100  $M_\odot$ 
Schwarzschild black hole 
may not have  a higher color temperature
than  $\sim$1.3 keV or  $\sim$0.7 keV, 
respectively, unless super-Eddington mass accretion  is allowed.
To explain higher  disk temperatures,
either super-Eddington mass accretion rates or  unreasonably small mass is required.
This is the ``too-hot disk'' problem 
 of the 
Galactic superluminal jet sources and ULXs
we are concerned in this paper (see also King and Puchnarewicz 2003).

\subsection{Multi-color disk approximation}

We point out that the maximum disk color temperature as expressed with
(\ref{Tcolmax}) is a quantity directly 
constrained from observations.  If we approximate
optically thick accretion disk spectra with a simple multicolor 
disk blackbody model (MCD model; Mitsuda et al.\ 1994; Makishima et al.\ 1996) in which 
local emission is a blackbody and radial dependence of the temperature is simplified as 
$T(r) = T_{in} \; (R_{in}/r)^{3/4}$,  the apparent 
inner disk radius $R_{in}$ and temperature $T_{in}$ 
will be the independent model 
parameters\footnote{In this paper, 
$r_{in}$ denotes the real disk inner radius, while $R_{in}$
is the apparent inner radius as a MCD model parameter.
Appendix \ref{GradDiskbb} gives conversion formulae between the MCD parameters
($R_{in}$ and $T_{in}$)
and realistic disk parameters ($M$ and $\dot M$)
in the Schwarzschild case.}, such that the spectral shape is determined only by $T_{in}$, 
and the flux is proportional to $R_{in}^2 T_{in}^4$.  
When observed spectra are fitted with the MCD model,
 $T_{in}$ may be identified with
the maximum disk color temperature (\ref{Tcolmax}),  while the
true inner disk radius $r_{in}$ and apparent radius 
$R_{in}$ have a relation
$r_{in} \approx 0.4 \; (T_{col}/T_{eff})^2 \: R_{in}$  
(Kubota et al.\ 1998).

\subsection{Local Energy Spectra}
\label{local}
In the inner part of the optically thick accretion disk, the electron scattering opacity
is dominant, and the hot disk atmosphere distorts the emergent spectra via Comptonization.
Precise theoretical
calculation of the   accretion disk spectra  for Galactic black hole candidates
has been made by several authors (e.g., Shimura and Takahara 1995; Ross and Fabian 1996; Blaes et al.\ 2001).  These authors tend to  agree that 
the local spectrum from each ring of the disk 
can be approximated
 by  the ``diluted blackbody'',  $(T_{eff}/T_{col})^4\:B(E, T_{col})$,
in  the $\sim$0.5 -- 10 keV band,
where $B(E, T_{col})$ is the Planck function with color temperature $T_{col}$ ($>T_{eff}$).
 Also,
 ratio of the color temperature to the effective  temperature, $T_{col}/T_{eff}$,  is 
virtually constant at 1.7 --  1.9  along the disk radius
for high values of the accretion  rate and low values of the viscosity parameter
(Shimura and Takahara 1995).
Hence,  as long as $T_{col}/T_{eff}$ is 
assumed  constant, 
equation  (\ref{Tcolmax}) is valid and simple MCD approximation may be used to
describe observed disk spectra.  
Throughout this paper, we assume $T_{col}/T_{eff}$ = 1.7 and
the diluted blackbody local spectra 
 (see discussion in  section \ref{LargeTcolTeff} for the limitation
of this assumption).

It should be noted that the mass obtained by fitting the standard accretion 
disk model to the observed spectra  is proportional to $(T_{col}/T_{eff})^2$, as long 
as the diluted blackbody approximation is valid.  This can be readily seen as follows: The accretion 
disk spectrum may be expressed as, 
$$
\int_{r_{in}}^{r_{out}} 2 \pi r  
\left(\frac{T_{eff}}{T_{col}}\right)^4\; B\left(E, T_{col}(r, M, \dot M)\right) \;dr
$$
$$
\approx {r_{in}}^2\int_{{(r/r_{in})}=1}^{\infty} 2 \pi (r/r_{in})
\left(\frac{T_{eff}}{T_{col}}\right)^4\; B\left(E, T_{col}(r/r_{in}, M, \dot M)\right) \;d(r/r_{in})
$$
\begin{equation}
= {\left(\frac{6G(T_{eff}/T_{col})^2M}{c^2}\right)}^2\int_{{(r/r_{in})}=1}^{\infty} 2 \pi (r/r_{in})
 B\left(E, T_{eff}(r/r_{in}, (T_{eff}/T_{col})^2 M, \dot M)\right) \;d(r/r_{in}), \label{MassTcolTeff}
\end{equation}
where we have used $r_{in}=6 GM/c^2$ and $T_{col}
\propto \dot M^{1/4} \left(r/r_{in}\right)^{-3/4} \left((T_{eff}/T_{col})^2 M\right)^{-1/2}$ from equation (\ref{Teff}).
Equation (\ref{MassTcolTeff}) indicates that the 
accretion disk spectrum with the black hole mass $M$ and the
diluted blackbody local spectrum is identical to the disk spectrum
with the black hole mass $ (T_{eff}/T_{col})^2 \: M$
and the blackbody local spectrum.

\subsection{Relativistic Effects}

\label{RelativisticEffects}

Due to the relativistic effects in the  Schwarzschild metric, 
observed disk spectral shape is mildly inclination-angle dependent,
such that inclined disk spectra become slightly harder (appendix \ref{GradDiskbb}, figure \ref{Schwarzschild_Kerr}). 
This  is  due to the gravitational redshift and Doppler boosts
near the inner edge of the accretion disk.  Note that the relativistic
effect is not very significant outside of $6\: r_g$, such that the 
Schwarzschild  disk spectra are not very much different from those 
of Newtonian disks.  This is in contrast to the extreme Kerr case,
in which  the  relativistic effect is enormous near $r_{in} \approx r_g$ and 
the disk spectrum is extremely inclination dependent (see below; figure \ref{Schwarzschild_Kerr}
and \ref{kerrdisk}).

\label{KerrDisk}
While in the Schwarzschild metric the last stable orbit
around the black hole is $ 6 \, r_g$,  in the
Kerr metric it can go down to $ 1.24\; r_g$
with an extreme angular momentum of $a=0.998$.
As the inner edge of the accretion disk  approaches the black hole, more gravitational
energy is released, and 
the inner disk temperature can get higher.
X-ray energy spectra from  optically thick Kerr accretion disks  have been calculated  by
many  authors including  Cunningham (1975), Connors, Piran and Stark (1980), 
Asaoka (1989), Sun and Malkan (1989),
Laor, Netzer and Piran (1990), Hubeny et al.\ (2000) 
and Gieli\'nski, Macio{\l}ek-Nied\'{z}wiecki  and Ebisawa (2001).  In this paper, we
calculate the Kerr disk spectral model  using the transfer function
by Laor, Netzer and Piran (1990) for $a=0.998$.
To make a comparison easier, we
assume $r_{in}=1.24\; r_g$, and  the same  local spectrum as we did for the Schwarzschild disk model
(section \ref{local}), 
namely diluted blackbody with  $T_{col}/T_{eff}$ = 1.7.

In Figure \ref{Schwarzschild_Kerr}, we show comparison of the Schwarzschild and Kerr
disk spectra, as well as Newtonian one,  for the face-on  and a highly inclined case ($i=80^\circ$).
The biggest difference between the Kerr disk model and the other two models is the presence or absence of the
innermost  region of the disk from $ 1.24\; r_g$ to $ 6 \, r_g$.
When the disk is close to face-on, the contribution from this
part is not significant because of  the gravitational red-shift and light-bending;
as a result, the total Kerr disk spectrum is not very much different from 
 the Newtonian  or Schwarzschild ones.
On the other hand, emission from the innermost part 
becomes very significant for a highly inclined disk due to Doppler boosting. 
Consequently,  a near edge-on  Kerr disk spectrum comprises a significant amount of 
high energy X-ray photons compared to the Newtonian or Schwarzschild disks.

Inclination angle dependence of  the observed Kerr disk spectrum  is  different 
in different energy bands (figure \ref{kerrdisk}).
At lower energies, where most of the emission is from the 
outer parts of the disk ($r > 400 \: r_g$), the flux is proportional to the
projected disk area and decreases with inclination, just as Newtonian disks.
 On the other hand, higher energy fluxes  from 
the innermost parts of the disk ($1.26\, r_g < r < 7 \, r_g$)
are  {\em enhanced}\/ as the disk is more inclined, due to strong
Doppler boosting. Consequently, inclined Kerr disks are
{\em brighter}\/ than the face-on disks in the highest energy bands, as previous
authors have already pointed out.

\section{Application of the Standard Disk Model}

We apply the standard accretion disk spectral
models to ASCA archival data  of GRO J1655--40 and IC342 Source 1.
The observations were made 
on 1995 August for GRO J1655-40 (when the source is in a  bright state) and 1993 September  for
IC342 Source 1 (when the source is in the ``high'' state; Kubota et al.\ 2001;
Kubota, Done and Makishima 2003).
These datasets may be considered ``exemplary'',
 so that 
the same spectral data of GRO J1655-40
have been analyzed  in Zhang et al.\ (1997), Zhang et al.\ (2000) and 
Gieli\'nski,    Macio{\l}ek-Nied\'{z}wiecki  and Ebisawa, K. \ (2001),
and those  of  IC 342 Source1 in Okada et al.\ (1998),
Mizuno et al.\ (1999), Watarai, Mizuno and Mineshige (2001) and 
Kubota et al.\ (2001).

We assume the diluted blackbody local spectra with $T_{col}/T_{eff}= 1.7$
in sections \ref{Standard1655} and \ref{ApplicationIC342}. Difference 
among the Newtonian, Schwarzschild, and Kerr disk models are only the
relativistic effects, including the variation of the innermost radius
and energy conversion efficiency (appendix \ref{efficiency_appendix}).
The Schwarzschild disk model we 
use (GRAD model) is explained in  Hanawa (1989) and  Ebisawa, Mitsuda and Hanawa (1991)
(see also appendix \ref{GRADbug}).  The GRAD model is available in the
XSPEC spectral fitting package, and this model also provides the  Newtonian disk spectra
we use in this paper,  when the relativistic
switch is  turned off.

\subsection{Application  to GRO J1655--40}
\label{Standard1655}
In Table \ref{GRO1655fit}, we show results of spectral fitting of the
ASCA GIS spectra
with Newtonian, Schwarzschild, and extreme Kerr disk models.
System parameters have been determined from optical observations
(Green,  Bailyn and Orosz 2003), such that
distance is 3.2 kpc, inclination angle is $70^\circ$,
and  $M=7 M_\odot$.
We fixed the distance, while several different inclination angles are tried from $0^{\circ}$ to
$80^{\circ}$.
The energy spectrum has  a power-law hard-tail  but contribution of which 
is minor below 10 keV and  hardly affects  discussion of the accretion disk spectrum.
The power-law slope is fixed at 2.5 which matches the simultaneous high energy observation 
with BATSE (Zhang et al.\ 1997), and its normalization is allowed to be free
within the range acceptable by  BATSE.
Hence, the free parameters are $M, \dot M$, power-law normalization, and the hydrogen column density.
Besides minor local features (see Gieli\'nski, Macio{\l}ek-Nied\'{z}wiecki and Ebisawa 2001),
all the models with different inclination angles can reasonably fit the observed 
accretion disk spectrum.

\begin{figure}
\centerline{
\includegraphics[angle=-90,width=0.45\columnwidth]{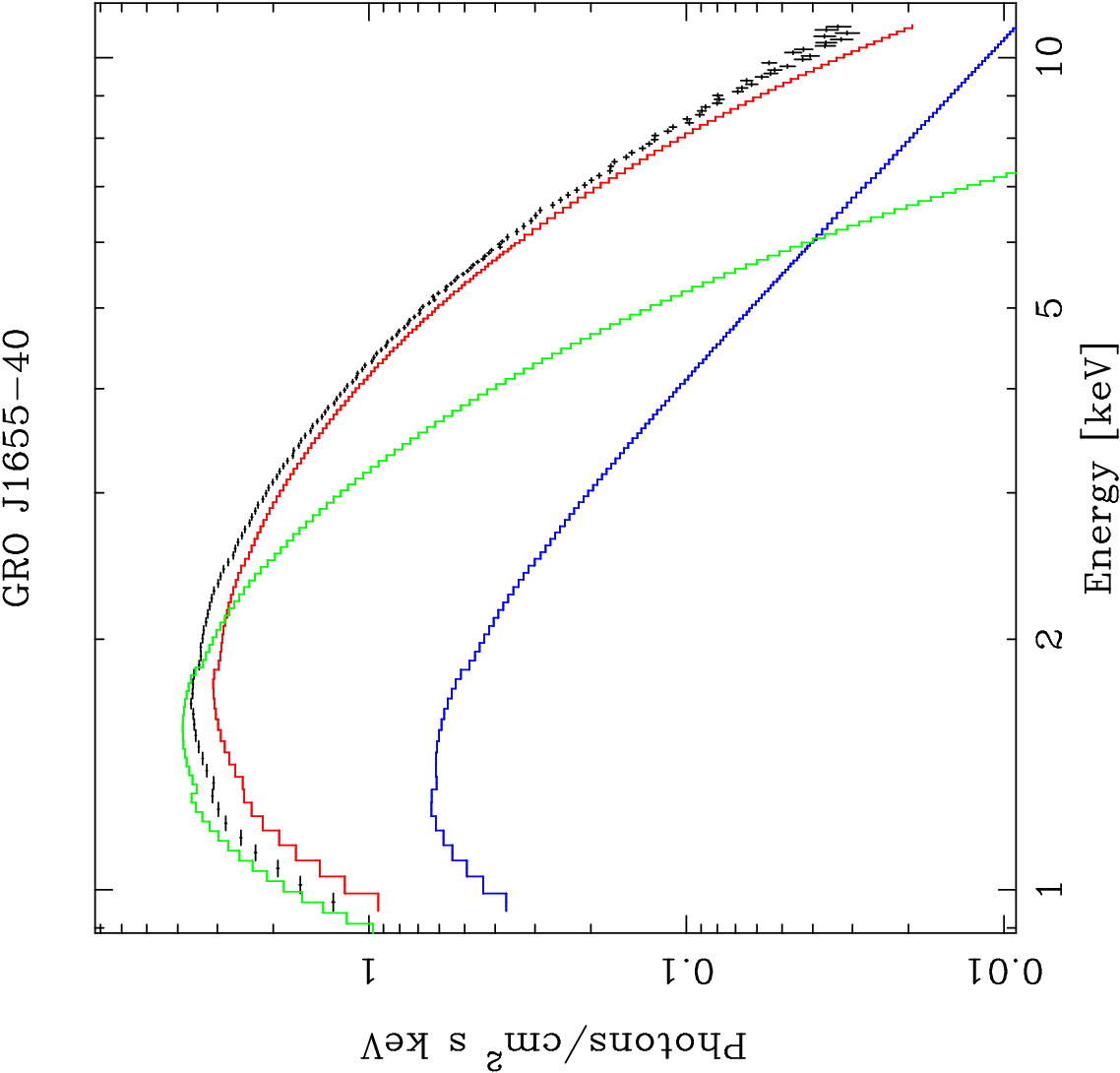} 
\hfil
\includegraphics[angle=-90,width=0.45\columnwidth]{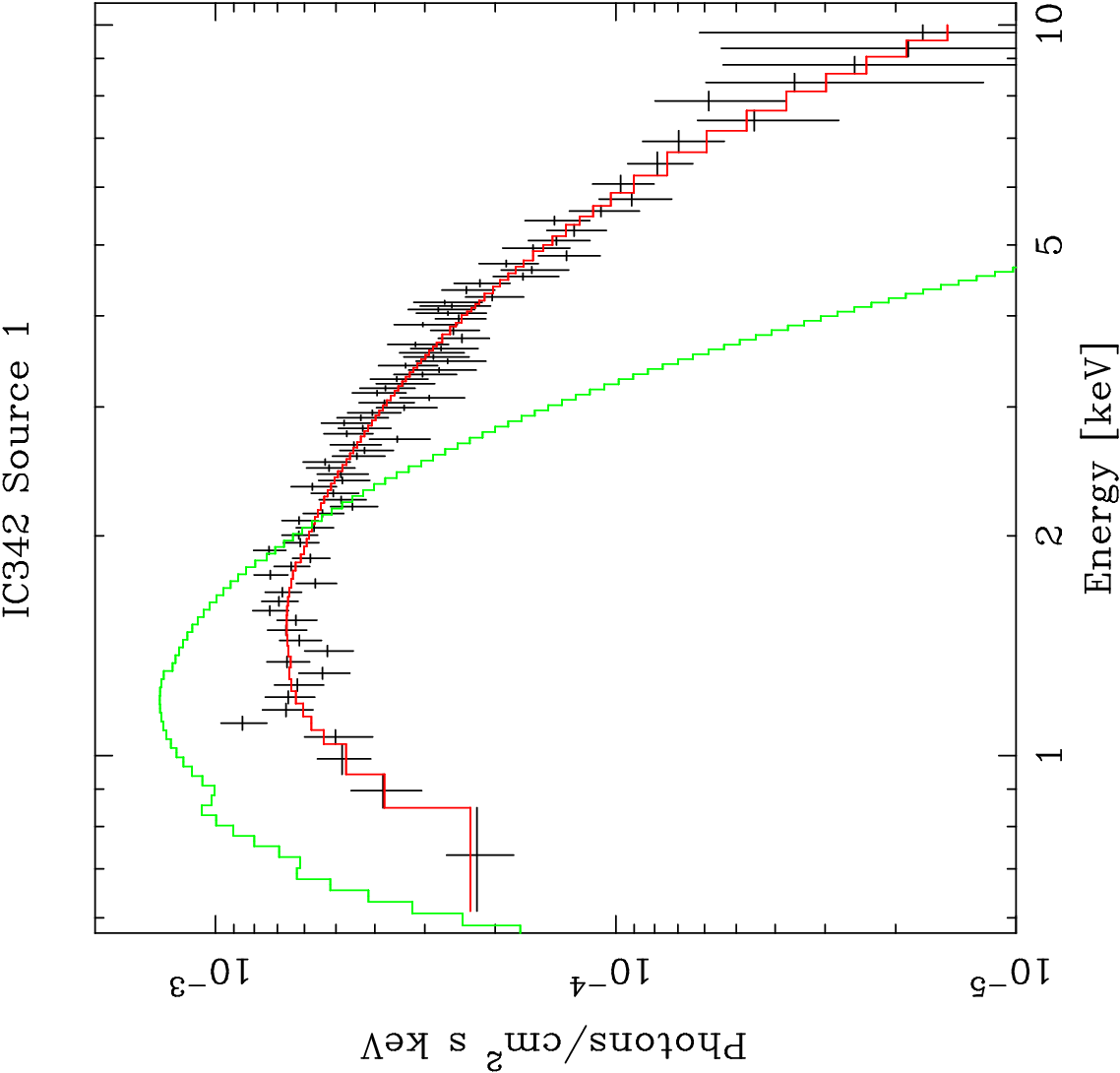}
}
\caption{
ASCA GIS energy spectra of  GRO J1655--40 (left panel) and 
IC342 Source 1 (right).
The best-fit Schwarzschild disk models are indicated in red,
in which case the best-fit  masses are too small 
(1.8  $M_\odot$  and 8.9$M_\odot$
for GRO J1655--40 and IC342  respectively)
compared to the mass determined from optical observations 
(7 $M_\odot$  for GRO J1655--40), or that
expected from the observed luminosity
(100 $M_\odot$  for IC342).
Schwarzschild disk spectra with
the expected masses give 
too low temperatures to explain the observed  accretion disk spectra for both cases
 (shown in green).
Additional power-law component (blue in left panel) is required to fit the  GRO J1655--40
spectrum.
}\label{Spectra}
\end{figure}

Note that the Schwarzschild disk model at the correct inclination angle
($i=70^\circ$) gives a too small mass ($1.8 M_\odot$) which is  unacceptable
(red curve in Figure \ref{Spectra} left).    If we fix the mass 
at the correct value ($7 M_\odot$), the disk spectrum cannot be hard enough  to explain
the observed spectrum (the ``best-fit'' is shown with green curve in Figure  \ref{Spectra} left):
The ``too-hot'' disk problem is  thus evident  (equation \ref{Tcolmax}).

It is of interest to see  how the black hole mass depends on the inclination angle.
Observation can constrain the projected disk area, which is proportional
to square of the mass.
Hence  $M \propto 1/\sqrt{\cos i}$ in  the Newtonian case, and the black hole mass at $i=80^\circ$ is
2.4 times higher than that for the  face-on disk. On the other hand, the mass increasing
factor is  2.9  in
the Schwarzschild case, and as much as $\sim$ 10 in the extreme Kerr case.
This is because inclined relativistic disk spectra become harder,
which is most conspicuous  in the extreme Kerr disk (section \ref{RelativisticEffects};
figure \ref{Schwarzschild_Kerr}).
Given the observed spectra, to compensate the spectral hardening with
inclination,  the mass becomes necessarily larger (equation \ref{Tcolmax}).

The extreme ($a$=0.998) Kerr disk model with the
correct inclination angle gives 
 $M = 15.9 M_\odot$.
Compared to the  fit with Schwarzschild disk model ($1.8 M_\odot$), 
significant increase of the mass does indicate the spectral
hardening of the Kerr disk model.
In fact,  the derived mass is too large
 to be consistent with the  realistic mass 7 $M_\odot$,
which probably suggests that 
$a$=0.998  is  too high.
In fact, Gieli\'nski, Macio{\l}ek-Nied\'{z}wiecki and Ebisawa (2001) applied the Kerr disk model to
the same ASCA energy spectrum of GRO J1655--40,
and concluded that  $a $ is likely to be between 0.68 and 0.88 to be consistent with
$M = 7 M_\odot$.  To summarize, in agreement with
Gieli\'nski, Macio{\l}ek-Nied\'{z}wiecki and Ebisawa (2001), a standard Kerr disk model may
explain the observed  accretion disk spectra of GRO J1655--40 at the 
inclination angle $i=70^\circ$.  The Kerr metric is required, but
not with the maximum  angular momentum.

That  GRO J1655--40 has probably a spinning black hole is also 
suggested by the recent discovery of 450 Hz QPO (Strohmayer 2001),
as this  frequency  is higher than the Kepler frequency at the
innermost stable orbit of a $7\, M_\odot$ Schwarzschild black hole.
Precise analysis of the QPO characteristics from GRO J1655--40 suggests that
the black hole is in  neither a Schwarzschild nor a
maximal Kerr black hole (Abramowicz and Klu\'zniak 2001),
which is consistent with our energy spectral analysis.

\subsection{Application  to IC342 Source 1}

\label{ApplicationIC342}

We assume the distance to the source 4 Mpc (Okada et al.\ 1998 and references there in).
If isotropic emission is assumed, the luminosity will be
$1.7 \times 10^{40}$ erg s$^{-1}$ (1 -- 10 keV) at this distance,
thus $M \gtrsim 100  M_\odot$ is expected so as not to
exceed the Eddington luminosity.
In table \ref{IC342fit}, we summarize results of the spectral fitting of the ASCA GIS spectrum 
 with Newtonian, Schwarzschild, and Kerr disk models, for different inclination angles.
All the fits are acceptable,  thus models may not be discriminated based on the
quality of the fits.
Note that the disk luminosities always exceed the Eddington limit more than
10 times in the Schwarzschild case.

The right panel in figure \ref{Spectra} indicates the  ASCA GIS spectrum
and the best-fit Schwarzschild disk model (in red).
For  comparison,   we show a Schwarzschild disk spectrum with
$M=100  M_\odot$ at the Eddington limit (in green);
we can clearly see that such an accretion disk has a too low temperature to 
explain the observed hard spectrum.

The face-on Kerr disk model
gives  $M = 27.3 M_\odot$ and $ \dot M =  2.0 \times 10^{20}$ g s$^{-1}$ ($ \dot M= 16 \; \dot M_C$).
A factor of $\sim$ 3 increase of the mass 
compared to the Schwarzschild case
is  due to slight   hardening of the face-on Kerr disk spectrum.  
Note that {\em the super-Eddington problem does not disappear},
as the face-on Kerr disk spectrum is not very different from the
Schwarzschild one (section \ref{RelativisticEffects};
figure \ref{Schwarzschild_Kerr}).
If we assume a very inclined disk with  $i=80^\circ$, we obtain 
$M = 332 M_\odot$ and $ \dot M =  1.5 \times 10^{20}$
g s$^{-1}$ ($\dot M = 1.0 \; \dot M_C$).
Now the super-Eddington problem is solved, that  is a consequence of the
fact that the inclined Kerr disk spectrum is  much harder  than the Schwarzschild one.
However, there will be two serious problems to accept the inclined Kerr disk model for 
ULXs in general:
First,  we do not know a mechanism to create   $\sim 350 M_\odot$ black holes.
Second, it is very unlikely that most of the accretion disks in ULXs are largely inclined 
when seen from the earth.

\subsection{Examination of Variants of Standard Disk Model}

In this section, we critically examine two  possibilities which have been
proposed to  make  the standard accretion disk spectra  look harder
and to explain the observed hard  spectra of ULXs and Galactic super-luminal jet sources.

\subsubsection{Large $T_{col}/T_{eff}$?}

\label{LargeTcolTeff}

 A naive solution to explain the apparently hard
 disk spectrum may be to allow $T_{col}/T_{eff}$ to be much
 greater than the standard value $\sim$ 1.7.
In this case  the color temperature of the disk can be much higher for the same mass 
and mass accretion rate, while
the effective temperature remains the same  (equation \ref{Tcolmax}).
In other words, for a given observed disk flux and spectrum, 
$M \propto  (T_{col}/T_{eff})^2$, thus  a larger mass is allowed
(equation \ref{MassTcolTeff}).

In fact, in the case of GRO J1655--40, 
(Section \ref{Standard1655}, 
table \ref{GRO1655fit}), if $T_{col}/T_{eff}=1.7 \sqrt{7 M_\odot/1.8 M_\odot} = 3.4$,
the observed spectrum is explained with the  Schwarzschild disk model 
with  $7 M_\odot$ for the same distance and inclination angle.
 Borozdin et al.\ (1998) reached a similar conclusion that   $T_{col}/T_{eff} = 2.6$ 
is required for GRO J1655--40 to fit with a Newtonian disk model whose inner disk radius is 
3 times the Schwarzschild radius. For IC342 Source 1, we need a still higher value of
 $T_{col}/T_{eff}=1.7 \sqrt{100 M_\odot/8.9 M_\odot}$ =5.7, to fit with a face-on
Schwarzschild disk model around a $100 M_\odot$ black hole.

In this manner,  extremely
high  $T_{col}/T_{eff}$ values might solve the ``too-hot'' disk
problem.  In fact, it is suggested that standard disks 
with high viscosity may require such  high 
values of the spectral hardening factor (Shimura and Takahara 1995).	
Given our lack of firm theoretical understanding of the 
mechanisms of viscous angular momentum transport (e.g., Merloni 2003), 
it would be better not to completely exclude the possibility of extremely large values of 
 $T_{col}/T_{eff}$ at this moment. 

On the other hand, there are several circumstantial evidences
that  $T_{col}/T_{eff}  $ cannot significantly exceed $\sim 2$:
Independent theoretical calculations agree with  values of  
$T_{col}/T_{eff}= $ 1.7 -- 1.9
for accretion disks around $\sim 10 M_\odot $black hole
for high values of the accretion  rate and low values of the viscosity parameter
(Shimura and Takahara 1995; Ross and Fabian 1996; Blaes et al.\ 2001).
Values of  $T_{col}/T_{eff}$ increase very slowly with  mass, 
but cannot be higher than $\sim 2.5$ even for a $10^6 M_\odot$
black hole
 (Ross, Fabian and Mineshige 1992; Blaes et al.\ 2001).
For observed accretion disk spectra of  Galactic black hole candidates or 
weakly magnetized neutron stars, 
$T_{col}/T_{eff}$ = 1.7 --1.9 gives reasonable  mass values which are consistent 
with those determined from  dynamical measurements
(e.g., Ebisawa, Mitsuda and Hanawa 1991; Ebisawa et al.\ 1993; Shimura and Takahara 1995;
Dotani et al.\ 1997).  

\subsubsection{Comptonization of disk photon?}

\label{Compton}
{\em If}\/ a  standard optically thick disk is
shrouded by a  thermal corona whose temperature
is slightly higher than the disk temperature, significant parts of the disk photons
are comptonized and appear in higher energy bands.
Thus, in such a situation, the observed hard energy
spectra from Galactic superluminal jet sources and
ULXs might be explained.


The comptonization model may phenomenologically fit the observed  spectra of
GRO J1655-40  (Zhang et al.\ 2000;   \.Zycki, Done and Smith 2001). 
Fitting with a thermal comptonization model (in which seed photons
have a single temperature blackbody spectrum)
gives typically the blackbody  temperate $\lesssim  0.5$ keV, 
plasma temperature $\sim $ 1.5 keV and  scattering optical depth $\sim $10.
At such a low temperature and high optical depth, the heavy elements
may not be fully ionized, and the plasma can be  optically thick to photoelectric
absorption.
Presumably, such a physical situation is more  reasonably
described by standard optically thick accretion disk in which
emergent spectrum is distorted by Comptonization ($T_{col} > T_{eff}$), 
which successfully explain the observation (section \ref{Standard1655}).

We  applied the comptonization model to IC342 Source 1 in 1993.
When  the seed photon is assumed to be disk blackbody, we obtain
$T_{in} \approx$  1.2 keV, plasma temperature $\sim$ 2.5 keV, and scattering optical depth
$\sim$ 10 ($\chi^2$/dof=51.8/46).
Note that the $T_{in}$ cannot go down to $\sim 0.7 $ keV as desired
for the Schwarzschild disk around a 100 $M_\odot$  black hole.
Therefore, the original ``too-hot'' disk problem may not be solved
even introducing the ad hoc comptonization plasma to harden the standard disk spectrum.
When the seed photon spectrum is assumed to be blackbody,
the blackbody temperature is  0.52 keV, the plasma temperature is
 1.8 keV  and the optical depth is $\sim $ 18 ($\chi^2$/dof=51.8/46).
Although this model may fit the data, it is hard to interpret these parameters
in the physical context, and the super-Eddington problem is unanswered. 
Presumably, the slim disk model explained in the
next section is more likely for ULXs when they are in the bright state.
Significant disk comptonization may be taking place, instead,
when ULXs are much dimmer and their energy spectra are power-law like
(Kubota, Done and Makishima 2003).

\section{The Slim Disk Model}

\subsection{Characteristics of  the slim disk model}
\label{slim-disk}
So far, we have considered the standard optically thick accretion disk 
(Shakura and Sunyaev 1973) in which radial energy advection
is neglected and all the gravitational energy
release is converted to thermal radiation.
In accretion disk theory
(for a review, e.g., Kato, Fukue and Mineshige 1998), 
there is another stable optically thick   solution,
which is called  ``optically thick ADAF (advection-dominated accretion flow)'' disk
or, ``slim'' disk (Abramowicz et al.\ 1988),
which takes place  when mass accretion rate is around the super-Eddington rate or higher.
In the slim disk,  all the gravitational energy release is not converted
to the thermal radiation,  but significant part of the energy is carried inward
 due to radial advection. Slim disk is geometrically thick, and can be much
hotter than the standard disk (Kato, Fukue and Mineshige 1998; Watarai et al.\ 2000).

One of the most significant observational characteristics of the
slim disk is that the disk luminosity can  exceed the Eddington limit
 by up to $\sim$ 10 times  (Abramowicz et al.\ 1988; 
Szuszkiewicz, Malkan and Abramowicz 1996; Kato, Fukue and Mineshige 1998).
This may be qualitatively understood as follows: At any disk radius, local  radiation 
pressure may not exceed the vertical gravitational force, such that
\begin{equation}
F(r) \lesssim \frac{c G M}{\kappa\,  r^2} \frac{h}{r},
\end{equation}
where $F(r)$ is the energy flux, $r$ the disk radius and $h$  the half-thickness.
Therefore,
$$
L_{disk} \equiv 2 \int_{r_{in}}^{r_{out}} 2 \pi r F(r) d r
$$
$$
\lesssim \frac{4 \pi c G M}{\kappa}\int_{r_{in}}^{r_{out}} \frac{h}{r^2} dr
$$
\begin{equation}
\approx L_{Edd}\; \left(\frac{h}{r}\right) \; \ln\left(\frac{r_{out}}{r_{in}}\right),\label{Ledd}
\end{equation}
where $r_{in}$ and $r_{out}$ are inner and outer disk radius, respectively,
such that $\ln\left(r_{out}/r_{in}\right) \approx 10$.
In the standard disk, which is geometrically thin, $h/r \lesssim 0.1$, thereby
$L_{disk} \lesssim L_{Edd}$.  On the other hand, with the slim disk
in which  $h/r \approx 1$, $L_{disk} \lesssim 10\: L_{Edd}$.

\begin{figure}
\begin{center}
\includegraphics[angle=-90,width=0.8\columnwidth]{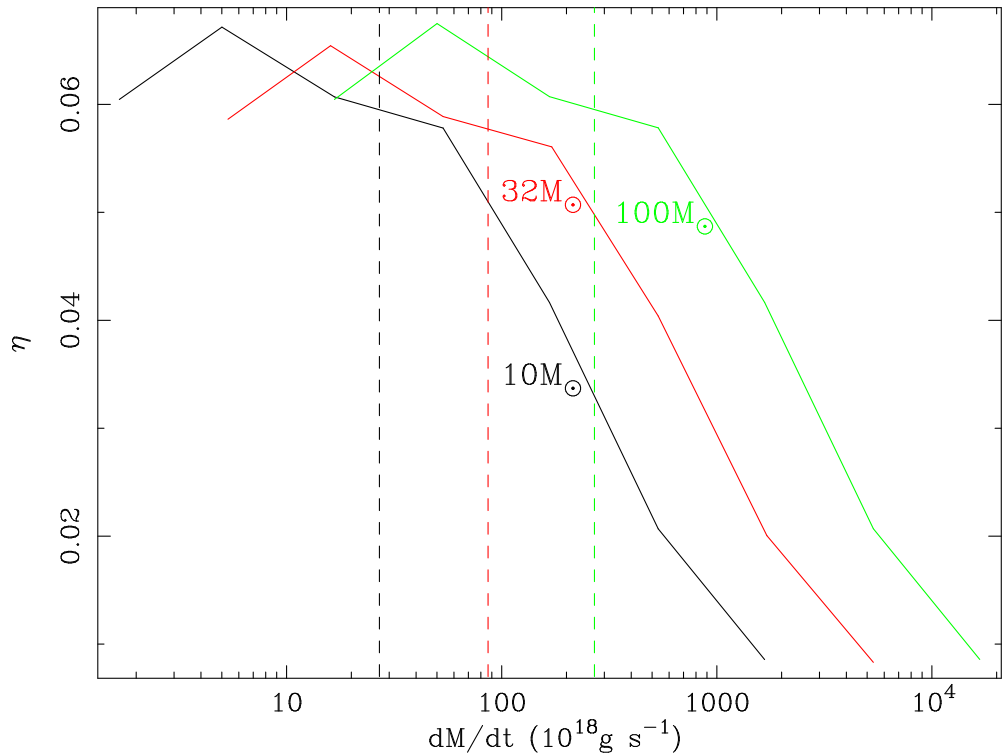}
\end{center}
\caption{Energy conversion efficiency of the slim disk used in the present paper and
in Watarai, Mizuno and Mineshige (2001). The vertical broken lines indicate
the critical mass accretion rates $\dot M_C$ which gives the Eddington luminosity
in the standard disk model
(see appendix \protect\ref{efficiency_appendix}).
}\label{SlimEfficiency}
\end{figure}

In addition to that super-Eddington luminosities are permissible, 
slim disk has following observational characteristics
 (e.g., Kato, Fukue and Mineshige 1998; Watarai et al.\ 2000; Watarai, Mizuno and Mineshige 2001):
(1) As the mass accretion rate exceeds the critical  rate $\dot M_C$, 
 energy conversion efficiency decreases due to advection
(appendix \ref{efficiency_appendix}, figure \ref{SlimEfficiency}).
Hence  the disk luminosity is no longer proportional to the mass accretion rate, but saturates at
$\sim$10 $L_{Edd}$.
(2) Innermost disk radius  can be closer to the black hole 
than the last stable orbit,  as  mass accretion rate increases. Even in the Schwarzschild metric, 
innermost disk region  inside $6\: r_g$ can emit significant amount of X-rays.
Thus slim disk
can be ``hotter'' than the standard disk.  
 (3) If radial dependence of the disk effective temperature
 is written as $ T_{eff} \propto r^{-p}$ (``$p$-free disk'' below), the exponent $p$ reduces from 0.75 
(which is expected for the standard disk) to 0.5 as advection
 progressively dominates. Thus spectral difference from the standard disk will be noticeable.

\subsection{Slim disk for Galactic Sources?}

\label{SlimDisk}
Watarai et al.\ (2000)  suggested that the 
slim disk model may explain
the apparently hard energy spectra of Galactic superluminal jet sources.
 In fact,  energy spectra of 
GRO J1655--40 (Kubota 2001),  GRS 1915+105 (Yamaoka 2001) and  XTE J1550-56 (Kubota 2001)
are not fitted with a standard
accretion disk model when their disk luminosities and temperatures reached
maxima, but  better fitted with the $p$-free disk model with  $p$ between 0.75 and 0.5.
This is considered to be an evidence of emergence of the slim disk,
when mass accretion rates are  extremely high in these sources.

On the other hand,  when the disk luminosity and
temperature are lower,  GRO J1655--40 is 
considered to embrace the standard disk, since the standard disk model can fit
the observed spectra well, and the inner disk radius is fairly constant 
over a large  luminosity variation (Sobczak et al.\ 1999;
Kubota, Makishima and Ebisawa 2001).
Without advection, the standard Kerr disk with high inclination 
is successful to explain the observed hard disk spectra (section \ref{Standard1655}).
Therefore, except when the disk luminosity is near the peak
and the slim disk presumably takes place, 
the standard disk is expected to be  present 
in the Galactic superluminal jet sources.

\subsection{Application to IC342 Source 1}

As  for ULXs, the slim disk model seems to be more likely,
as ULXs  are intrinsically bright systems.
Mizuno, Kubota and Makishima  (2001) studied spectral variations of
several ULXs, and commonly found anti-correlation between
the MCD parameters $R_{in}$ and $T_{in}$, which is another
observational feature of the slim disk (Watarai et al.\ 2000).
Watarai, Mizuno and Mineshige (2001)  calculated
X-ray energy spectra of slim disks, and  suggested that ULXs are slim disks 
around 10 -- 30 $M_\odot$ black holes shining more luminously  than 
Eddington limits.  A characteristic spectral transition has been observed in
IC342 Source 1, such that the bright state in 1993 was represented with the
MCD model spectrum, while in 2000 the source was dimmer and the energy spectrum was power-law like
(Kubota et al.\ 2001).  This spectral transition may be interpreted as
the disk state transition between 
the blight slim disk state  and another anomalous state  (Kubota,  Done  and Makishima 2002).

In the present  paper, we fit the observed energy spectra of IC 342 Source 1 in 1993
with the same slim disk model calculated by  Watarai, Mizuno and Mineshige (2001).
Namely, $\alpha=0.01$, and the local emission is a diluted blackbody with
$T_{col}/T_{eff}=1.7$. 
The model assumes the face-on geometry, 
and the distance is 4 Mpc.
The energy spectra are calculated for 
each grid point of $M$ = 10, 32, 100 $M_\odot$ and $\dot M/(L_{Edd}/c^2)=$
1, 3, 10, 32, 100, 320, 1000.  We fitted the observed spectra using XSPEC spectral fitting
package with this grid-model, so that XSPEC interpolates the model spectra for
other $M$ and $\dot M$ values.

Fitting result for IC342 Source 1 ASCA GIS spectrum is shown 
in table \ref{IC342fitslim}.
We find  $M =19.8  M_\odot$ and  $\dot M / \dot M_C \approx 20$.
This  mass accretion rate would give 20 times the Eddington luminosity
in the case of the standard disk.  However, since slim disk is less efficient at $\dot M \gtrsim \dot M_C$
(appendix \ref{efficiency_appendix} and figure \ref{SlimEfficiency}), the disk luminosity
is in fact only $\sim 6$  times the Eddington luminosity, 
which is  allowed in the slim disk (section  \ref{SlimDisk}).

\begin{figure}
\begin{center}
\includegraphics[angle=-90,width=0.65\columnwidth]{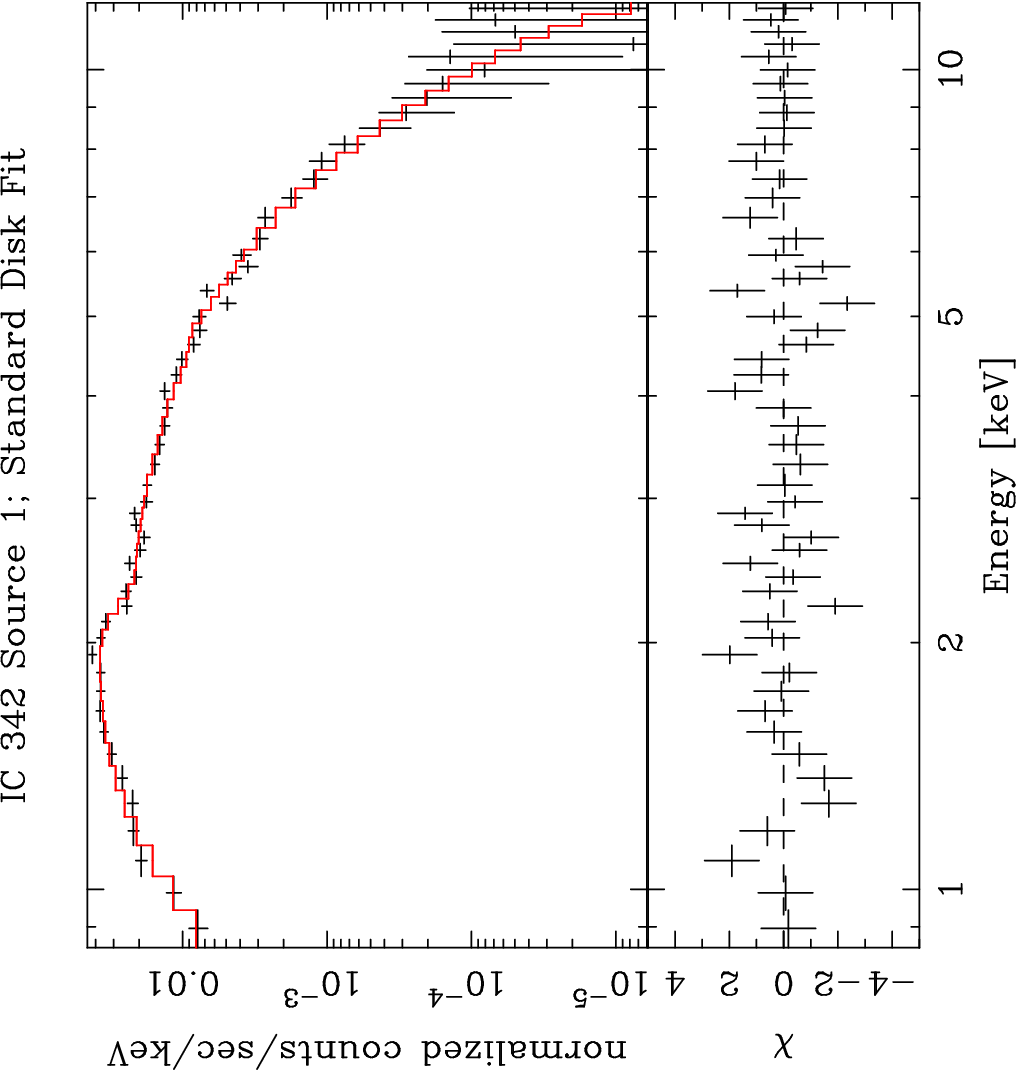}
\end{center}
\begin{center}
\includegraphics[angle=-90,width=0.65\columnwidth]{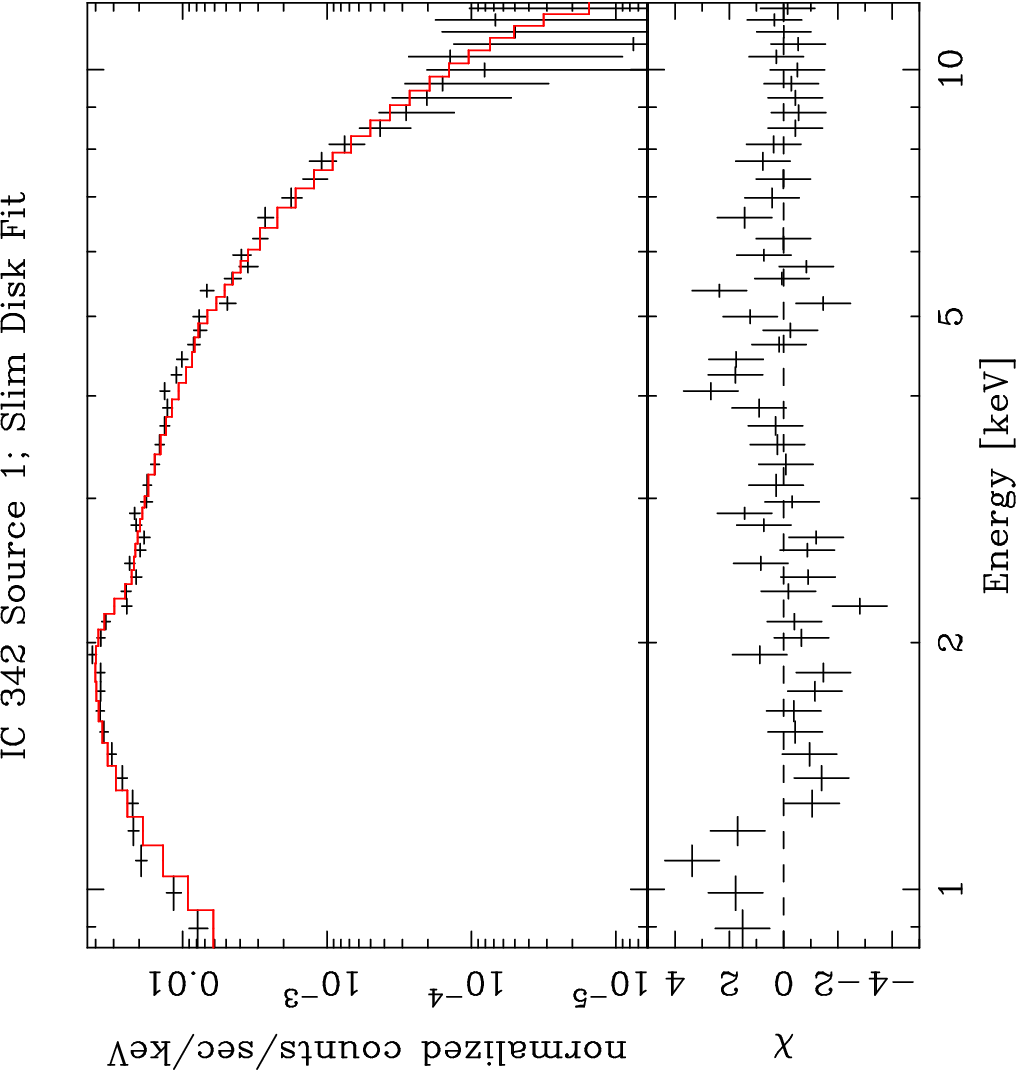}
\end{center}
\caption{IC342 Source 1 ASCA GIS spectrum fitted
with the Schwarzschild disk model ($i=0^\circ$; top)
and with the slim disk model (bottom).  Residual
is more conspicuous in the slim disk model below $\sim$ 1.5 keV.
}\label{IC342FitFigure}
\end{figure}

\begin{figure}
\begin{center}
\includegraphics[angle=-90,width=0.95\columnwidth]{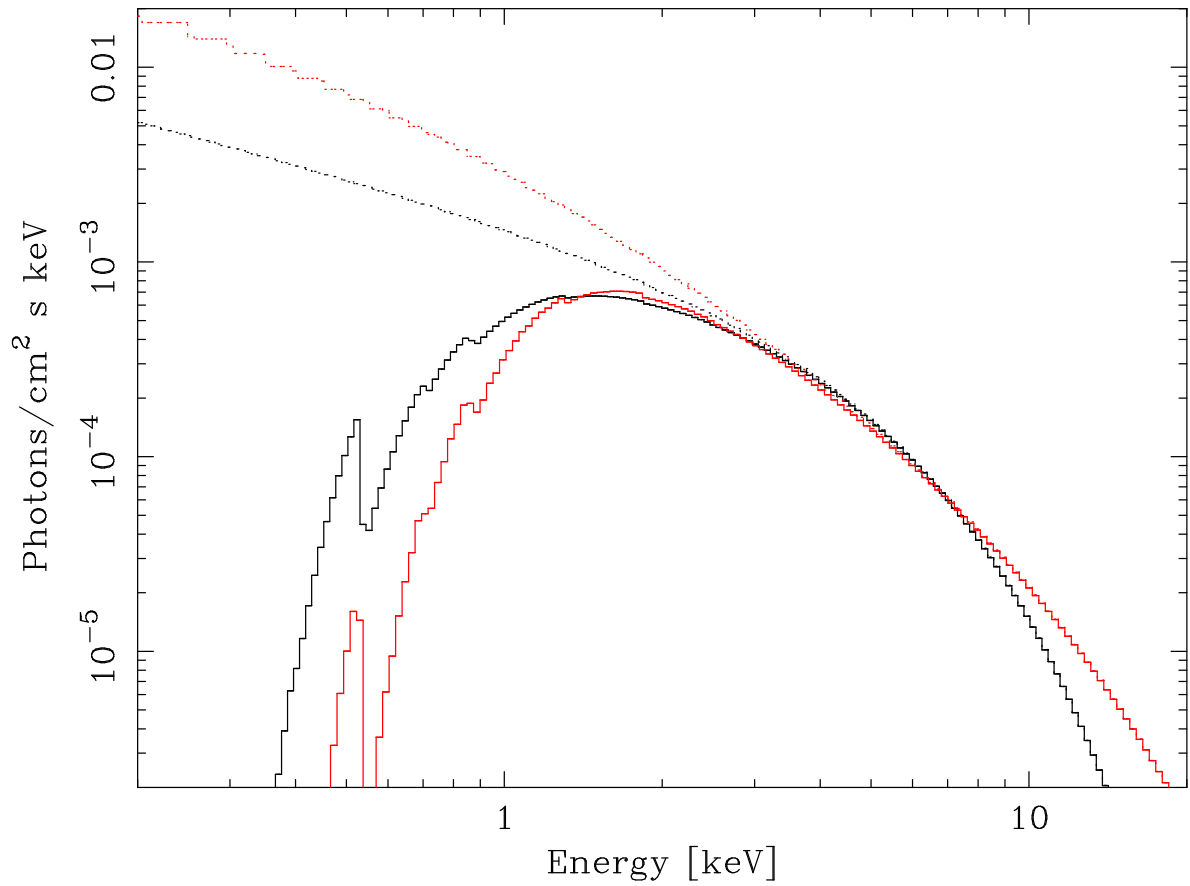}
\end{center}
\caption{IC342 Source 1 best-fit standard accretion disk model
(face-on Schwarzschild disk; black) and slim disk model (red). 
Models removed of interstellar absorption are shown in dotted lines.
}\label{IC342StandardSlim}
\end{figure}

The slim disk model can fit the observed spectra reasonably
well ($\chi^2$/dof = 1.40), but not as good as the
standard disk models ($\chi^2$/dof $\approx$ 0.90; table \ref{IC342fit})
(figure \ref{IC342FitFigure}).  Also,
the hydrogen column density is significantly larger with the
slim disk model ($\sim 9 \times 10^{21}$ cm$^{-2}$) 
compared to those with the standard disk models (3--5 $\times
10^{21}$ cm$^{-2}$).  This is because the spectral difference between the
best-fit slim disk model and standard disk spectra is largest in the
lower energy band (figure \ref{IC342StandardSlim}).
We simply assumed  diluted blackbody local spectra
with constant $T_{col}/T_{eff}$, which is known to be a good 
approximation for the standard disk (section \ref{local}).
However,  precise theoretical calculation of the local spectrum in the slim does not exist yet,
and  more realistic slim disk spectral model is expected to fit the observed spectra better
(section \ref{future}).

\subsection{Spectral Variation of  IC342 Source 1}
The characteristic anti-correlation between $R_{in}$ and $T_{in}$ 
discovered by Mizuno, Kubota and Makishima (2001) is 
considered to be  a characteristic of the slim disk.
Watarai, Mizuno and Mineshige (2001), 
through indirect comparison of the MCD parameters and
slim disk spectra,  claimed that this
spectral variation  may be  explained with  a slim disk at
a constant $M$ only by varying the mass accretion rate.
We demonstrate this claim by directly fitting the ASCA IC 342 Source 1 spectra
with the slim disk model.

\begin{figure}
\begin{center}
\includegraphics[angle=-90,width=0.75\columnwidth]{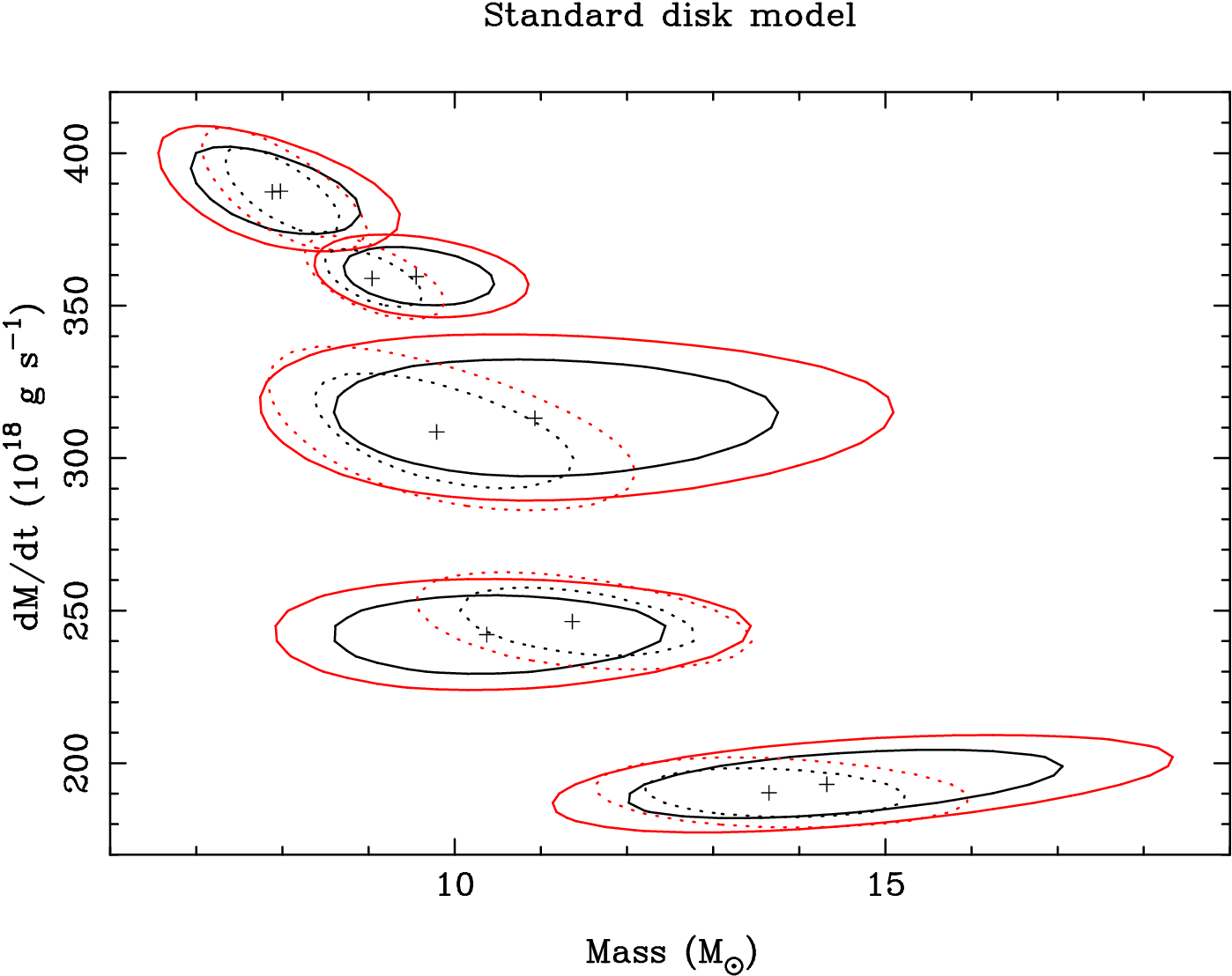}
\includegraphics[angle=-90,width=0.75\columnwidth]{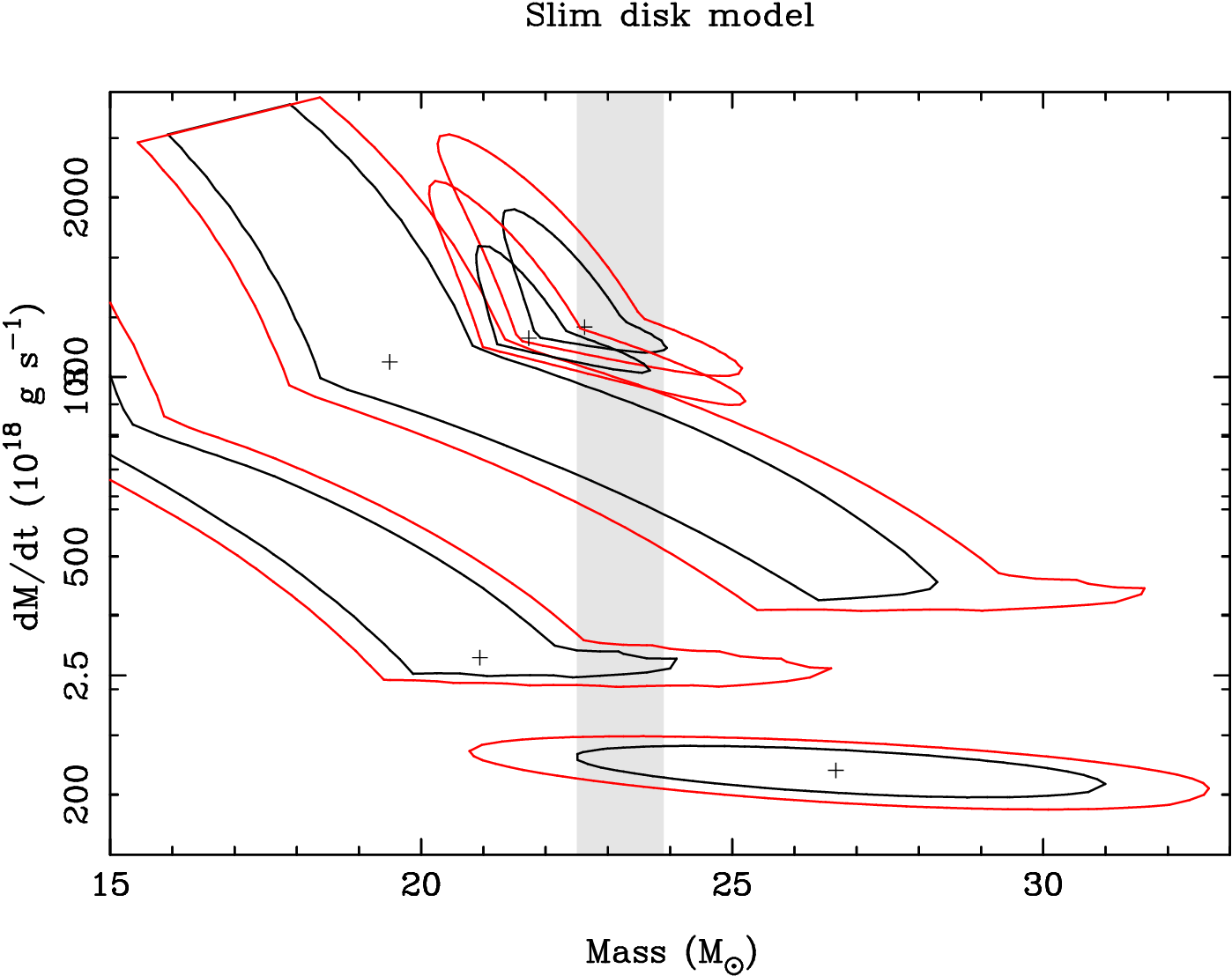}
\end{center}
\caption{Spectral variation of IC342 Source 1 observe in September 1993. The same datasets
as in Mizuno, Kubota and Makishima (2001) are used.  The upper-panel indicates the
variation of mass and mass accretion rates obtained by applying the standard
Schwarzschild disk model (GRAD model).  Contours with  dotted lines are from fitting in 0.5--10 keV,
and those with solid lines are from 2--10 keV. The two contour levels indicate  68 \% (1 $\sigma$) and
90 \% error regions for two parameters.
The lower panel is obtained through application of the slim disk model in 2 -- 10 keV.
The region marked with gray indicates that the observed spectral
variation is achieved with constant $M$ and variable mass accretion rates.
}\label{contour}
\end{figure}

We use the same spectral datasets used by Mizuno, Kubota and Makishima (2001).
The observation period is split into five periods depending on the
flux levels.  For each period, GIS and SIS spectra are fitted simultaneously to
achieve the better statistics.  In the top panel of figure \ref{contour},
we show $M$ and $\dot M$ variation obtained with the  face-on GRAD model.
We fixed the hydrogen column density at the average value ($N_H=5.2\times10^{21}$cm$^{-2}$), 
so that the free parameters
are only $M$ and $\dot M$. The reduced $\chi^2$ is less than unity for all the five spectra.
We see clear anti-correlation between these two parameters, which is just rephrasing the
$R_{in}$--$T_{in}$ anti-correlation discovered by Mizuno, Kubota and Makishima (2001).
Since $M$ must not  vary in the real world, this relation is telling that the standard accretion disk 
is not a proper model for IC 342 Source 1.  The spectral variation is clearly
seen  using only the  2 -- 10 keV band, though less clear than using the entire  0.5 -- 10 keV band.

Since the present slim disk model does not fit the observed spectra well
below $\sim $1.5 keV (table \ref{IC342fitslim} and figure \ref{IC342FitFigure}), 
we study spectral variation only using 2 -- 10 keV (then reduced $\chi^2$ will be $\sim$1).
For fair comparison, the hydrogen column density was fixed to the same 
average value as used in the GRAD model fitting.  The $M$ and $\dot M$ relation
obtained from the slim disk fitting is shown in the bottom panel in figure \ref{contour}.
We see that  $M$ in the range of  22.5 $M_\odot$ to 23.9$ M_\odot $
 is consistent with
all the five spectra.  Therefore, based
on the slim disk model,
we may interpret the observed characteristic spectral
variation of IC 342 Source 1 as a result of simple mass accretion rate variations.

\subsection{Future Problems of the Slim Disk Model}

\label{future}

Compared to the standard disk model,  X-ray spectral study  from the slim disk has 
only a short  history.  As already stated above, the present slim disk model
may not fit the observed ASCA spectra perfectly well, though the spectral
variation is better described with the slim disk model than with the standard disk.
There are several difficult problems  to  calculate the slim disk spectra 
precisely.
Although the  face-on geometry is assumed in the present slim disk model, 
observed flux from a slim disk is considered to be strongly inclination angle dependent.
 The flux  drops with inclination angle $i$ more rapidly than
$\cos i$, because the horizontal photon diffusion time-scale gets longer than the
in-fall time-scale (Kato, Fukue and Mineshige 1998).
Also, relativistic effects have yet to be taken into account, and 
it is pointed out that  
the disk around a Kerr black hole can be much hotter than the Schwarzschild case (Beloborodov 1998).
In addition, effects of radial photon-trapping may not be negligible under  extremely
high accretion rates (Ohsuga et al.\ 2002).  Due to these effects, 
$T_{col}/T_{eff}$ can be much higher than $\sim$1.7 
and may have a significant radial dependence (Shimura and Manmoto 2003;
Kawaguchi \ 2003).
More precise spectral model calculation 
from the slim disk is anticipated overcoming these difficulties.


\section{Origin of the  Ultra-Luminous X-ray Sources}

We have shown that the present slim disk model, though still primitive,  is likely to
explain the observed super-Eddington luminosities,  hard energy spectra,
and spectral variations of IC342 Source 1, in agreement with Watarai, Mizuno 
and Mineshige (2001).  Whereas
Watarai, Mizuno and Mineshige (2001) indirectly  compared the
observed spectra and the slim disk model spectra, we directly fitted the
observed spectra with the slim disk model, and made a quantitative comparison.
Thereby, we obtained the black hole mass  $\sim 20 M_\odot$, and the disk luminosity
$\sim 6$ times the Eddington luminosity for IC342 Source 1.  Thus, 
 ``intermediate mass
black holes'' are not required to explain the observed luminosity  $\sim 10^{40}$
erg s$^{-1}$.

In our Galaxy, the most massive stellar black
hole is $\sim 14 M_\odot$ in GRS 1915+40 (Greiner, Cuby and  McCaughrean 2001), 
as far as currently measured.  However, more massive stellar black holes
may well exist in other galaxies.  In fact,  stellar evolution theory says
 main-sequence stars can have a maximum mass   $ \sim 60 M_\odot$
(Schwarzschild and H\"arm 1959), and
the final black hole mass could be theoretically as large as its progenitor 
 beyond 40 $M_\odot$  (Fryer 1999). 
Observationally, Grimm, Gilfanov and Sunyaev (2003)
found  that there exists such a   universal luminosity function
of  X-ray binaries that is applicable to several different galaxies including  Milkyway,
where normalization for different galaxies are proportional to  the star-forming rates.
Galaxies  which are active in star formation tend to have more luminous and massive 
compact objects, and  the maximum cut-off luminosity is $\sim 10^{40}$ erg s$^{-1}$.
We suggest that  ULXs with luminosities of $ \sim 10^{40}$ erg s$^{-1}$ are 
black holes having a few tens of solar mass, which reside at the 
brightest end of the X-ray binary luminosity function.
Observed evidence of the association  between some  ULXs and star forming region may support our idea 
(e.g., Matsushita et al.\ 2000).

\section{Conclusion}

\begin{enumerate}
\item   In order to solve  the ``too-hot disk'' problem of ULXs and Galactic superluminal
jet sources, we have carefully studied energy spectra of the standard accretion disk and
slim disk.  In particular, 
we have calculated the extreme Kerr disk spectral model, and studied how
relativistic effects and disk inclination angle affect the observed disk spectra.

\item We have found that the standard Kerr disk model can successfully explain the
observe hard spectra of GRO J1655--40,
because the Kerr disk spectra become significantly harder than Schwarzschild 
ones when the disk is significantly inclined, which is exactly the case for 
  GRO J1655-40.  Another super-luminal jet source GRS 1915+105 is also
a highly inclined system (Greiner, Cuby and  McCaughrean 2001), so that  
the idea of inclined Kerr disk model may  be  plausible too.

\item The Kerr disk spectra are not significantly harder than the 
Schwarzschild disk when disk inclination angle is not large.
We conclude that the standard Kerr disk model is not appropriate 
to explain the observed hard spectra of most ULXs, since (1)  it is  unlikely
that accretion disks in most ULXs are preferentially inclined, and
(2)
if the inclined Kerr disk is applied, unreasonably large
black hole mass ($\sim 300 \; M_\odot$) is required.

\item We have calculated the slim disk spectra, and
applied to the observed spectra of IC342 Source 1 in 1993,
when the source is in the high state.
Although the current primitive slim disk model does not
perfectly fit the data, we found the slim
disk can explain the observed super-Eddington luminosity,
hard X-ray spectra, and spectral variation successfully.
In particular, the observed characteristic spectral
variation is explained with a constant mass at $M \sim 20 M_\odot$,
only varying the  mass accretion rate.

\item We propose ULXs are binary systems with a few tens of solar mass black holes,
which reside at the bright end of the X-ray binary luminosity function.
Such moderately massive stellar black holes may not exist in our Galaxy,
but presumably not uncommon in the galaxies where massive star formation is much
more active.

\end{enumerate}




\acknowledgments

We are grateful to Prof.\ K. Makishima and Dr.\ Luigi Foschini
for useful comments and discussion, 
and  to  Drs.\ D. Bhattacharya, 
S. Bhattacharya and A. V.   Thampan for pointing out  bugs in the old GRAD code 
and comparing the fixed GRAD code with their accretion disk model.
We are glad to Dr.\ Hanawa for developing the original GRAD code
and fixing the bugs.  
We acknowledge Dr.\ A. Laor for
making his transfer function calculation available with the XSPEC package.
We are thankful to the 
anonymous referee for many variable comments to improve the paper.
This research has made use of public data and software obtained 
from the High Energy Astrophysics Science Archive Research Center (HEASARC), 
provided by NASA's Goddard Space Flight Center. 
This work was supported in part by Polish KBN grant PBZ-KBN-054/P03/2001.



\appendix

\section{Energy Conversion Efficiency and Critical Mass Accretion Rates}
Energy conversion efficiency of optically thick accretion disk
($\eta$)\label{efficiency_appendix}
is dependent on  disk model assumptions,
so that a particular care is needed to interpret the disk parameters  we obtain from 
spectral model fitting.
In particular,  conversion efficiency of the  slim disk model is dependent on
the mass and mass accretion rates (Watarai et al.\ 2000; Watarai, Mizuno and Mineshige 2001).

In the Newtonian case, energy loss in the standard disk per second ($\equiv$ disk luminosity,
$L_{disk}$) may
be written as
$$L_{disk} = \dot M \left(\frac{GM}{r_{in}} - \frac{L^2}{2{r_{in}}^2}\right)$$
$$=\frac{GM\dot M}{2r_{in}},$$
where $r_{in}$ is
the innermost radius, and 
$L$ is the specific angular momentum at $r_{in}$, $\sqrt{GMr_{in}}$.  If we 
take $r_{in}$ as the last stable orbit 
in the Schwarzschild metric, $6\: r_g = 6 \;GM/c^2$, $L_{disk}=\frac{1}{12}\dot M c^2$, hence
$\eta=1/12$ for the Newtonian disk.  
In the Pseudo-Newtonian potential, which the present slim disk model adopts,
$$L_{disk} = \dot M \left(\frac{GM}{r_{in}-2r_g} - \frac{L^2}{2{r_{in}}^2}\right),$$
where $L =\sqrt{GM{r_{in}}^3/(r_{in}-2r_g)^2}$.
With  $r_{in}=6\: r_g$, we obtain $\eta$=1/16.  This is close enough
to the correct  efficiency in the Schwarzschild metric, $\eta$=0.057.
For the extreme Kerr case with $a=0.998$ and  $r_{in}=1.24\: r_g$, $\eta$  = 0.366.

We may define the critical accretion rate  $\dot M_{C}$ 
so that $L_{Edd} \equiv  \eta \dot M_C c^2  $, where  $L_{Edd}$ is  the Eddington luminosity. 
For the Schwarzschild case with $r_{in}=6\: r_g$,  $\dot M_{C} =
2.9 \times 10^{18} (M/M_\odot)$ g s$^{-1}$, and   
for the Newtonian case with the same innermost radius, $\dot M_{C} = 2.0 \times 10^{18} (M/M_\odot)$ g s$^{-1}$.
For the extreme Kerr case  $\dot M_{C} = 4.6 \times 10^{17} (M/M_\odot)$ g s$^{-1}$.

In the pseudo-Newtonian potential, if the innermost radius is constant 
at $r_{in}=6\: r_g$,  $\dot M_{C} = 2.7 \times 10^{18} (M/M_\odot)$ g s$^{-1}$.
However, in the present slim disk model, the inner radius becomes smaller than 
$6\: r_g $, and
 the energy advection is dominant at high accretion rates
(Watarai et al.\ 2000).
 Hence, the conversion efficiency is dependent on the mass and mass accretion rate.
In figure \ref{SlimEfficiency}, we show conversion efficiency of the slim
disk for several mass and mass accretion rate values. We see that at low 
mass accretion rate limits, $\eta \approx 1/16$, but the efficiency significantly decreases
with mass accretion rates when $\dot M \gtrsim \dot M_C$ due to the advection effect.

\section{Bugs in the original GRAD model}
\label{GRADbug}

Three mistakes have been found in the appendix of Ebisawa, Mitsuda and Hanawa (1991; EMH)
besides  obvious typos:
\begin{itemize}
\item (A8) in EMH should be read as follows:
\begin{equation}
\frac{d \theta'_{\rm ph}}{d \theta_{\rm ph}} = 
\frac{\cos i \sin \theta_{\rm ph} + \sin i \cos \thetaph \sin \phiph}{\sin \thetaph' }.\label{a8}
\end{equation}

\item (A15) in EMH should be read as follows:

\begin{equation}
\frac{\epsilon_{\rm DO}}{\epsilon_{\rm LO}} = \sqrt{1 - \frac{3GM}{rc^2}}
\left[
1 - \left(1-\frac{2GM}{rc^2}\right)^{- {\bf 0.5}}\sqrt{\frac{GM}{rc^2}}\frac{r}{c}
\frac{d\varphi'_{\rm ph}}{dt_{\rm ph}}
\right]^{-1}.\label{a15}
\end{equation}
The bold-face part was 1 in EMH, which is wrong.

\item (A22) in EMH  should be read as follows:
\begin{equation}
T_{\rm col} = 1.11\left(\frac{T_{\rm col}/T_{\rm eff}}{1.5}\right)
\left[\frac{g(r/r_g)}{0.15}\right]
\left(\frac{M}{{\bf 1.0}\:M_\odot}\right)^{-1/2}
\left(\frac{\dot M}{\rm 10^{18}\; g\; s^{-1}}\right)^{1/4}
 \;\; {\rm keV}.\label{a22}
\end{equation}
The bold-face part was 1.4 in EMH, which is incorrect.
\end{itemize}

Below,  equations (A1) to  (A4) refer to those in  the present appendix,
not in EMH.
Equation (\ref{a8}) was already fixed in the GRAD model code  used 
in EMH, which was distributed with XSPEC version 11.0.1ae and previous
ones.
Equations (\ref{a15}) and (\ref{a22}) were {\it not}\/ fixed in the code,
and the wrong values were used in EMH.  
In addition, there was another bug in the code used in EMH:   In the part to calculate the disk flux,
$$
      {\rm F = 8.9038633D-3*F/Ratio**4/(D/10.0)**2*(Em/1.4)**2}
$$
was wrong, and this has to be
\begin{equation}
      {\rm F = 8.9038633D-3*F/Ratio**4/(D/10.0)**2*{\bf Em}**2},\label{F}
\end{equation}
where Em denotes the compact object mass.
These bugs in equation (\ref{a22}) and (\ref{F}) compensate
each other (but not perfectly), which  probably explains why we could  not
notice these bugs much earlier.  The bug in equation (\ref{a15}) affects only 
when the disk is inclined, and this  effect is not significant.

Besides the bug in equation (\ref{a15}), one can see 
from equations (\ref{a22}) and (\ref{F}) that
the old GRAD spectra with $M=1.4 M_\odot$ and the 
new GRAD spectra with $M=1.0 M_\odot$ are identical.
Therefore, besides the  minor difference of redshifts which
originates in (\ref{a15}),
when  an  observed spectrum was fitted with the old GRAD model, 
{\em the mass of the compact object
was erroneously  1.4 times higher  than the correct mass given by the fixed GRAD model}.
This should be taken into account to interpret published results obtained using the old GRAD model
(e.g., Ebisawa, Mitsuda and Hanawa 1991; Kitamoto, Tsunemi and Roussel-Dupre 1992;
Ebisawa et al.\ 1993; Dotani et al.\ 1997; Kubota
et al.\ 1998).
The mass accretion rate represented in the dimension of mass per time is unchanged.
The fixed GRAD code is distributed with XSPEC version 11.1.0 and on.

Note that the bug in the old GRAD code may compensate with an ambiguity  of
$T_{col}/T_{eff}$ values.  In fact, in Ebisawa, Mitsuda and Hanawa (1991)
 and Ebisawa et al.\ (1993), we used $T_{col}/T_{eff} = 1.5 $ and
obtained reasonable masses of black holes and neutron stars, which should
be reduced by 1.4 times using the correct GRAD model. On the other hand,
the mass estimate is proportional to $(T_{col}/T_{eff})^2$ (when distance is known) or 
$(T_{col}/T_{eff})^4$ (when $d/\sqrt{M}$ is known), thus using $T_{col}/T_{eff}=$  1.8 or
1.6 instead of 1.5 retrieves the original mass values.  
The higher  $T_{col}/T_{eff}$ values are  in fact consistent with a 
precise calculation carried out more recently (Shimura and Takahara 1995;
Ross and Fabian 1996).

We confirmed that the fixed GRAD model gives almost identical
spectra with 
the Schwarzschild disk spectra calculated by  Bhattacharyya, Bhattacharya and Thampan (2001).
   Only a minor difference is found  when
the accretion disk is more inclined than $\sim 60^\circ$, when
the model by Bhattacharyya, Bhattacharya and Thampan (2001) gives as much as
$\sim 10$ \% higher flux at $\sim$ 10 keV.
The discrepancy is probably due to the fact that
the GRAD model does not take account of the photons 
whose trajectories change directions more than 180 degrees.  
In fact, those photons  may not be observable being 
blocked by the putative plasma near
the black hole which is responsible for  generation of the power-law hard-tail
component.

\section{Comparison of Newtonian and Schwarzschild Accretion Disk Spectra}
\label{GradDiskbb}
Comparison of the GRAD and MCD model parameters  have been made by EMH, which
is affected by the bug explained above.
Here, we present correct results using the new GRAD code after the bug  is fixed.
Also, we discuss how the relativistic effects 
(in the Schwarzschild case)
affect the mass and mass accretion rate determination from
observed accretion disk spectra.

MCD model does not take into account the inner boundary condition,
but  simply assumes $T(r) = T_{in} \; (R_{in}/r)^{3/4}$, where $R_{in}$
and $T_{in}$ are the {\em apparent}\/ inner disk radius and temperature,
respectively. MCD model is useful to represent observed spectra, since it
has the two independent parameters  which are directly constrained from
observation,
such that the spectral shape is determined only by $T_{in}$, and the normalization
is proportional to $R_{in}^2 \cos i$, where $i$ is the inclination.  Often disk inclination
is unknown, so that $R_{in}$ is uncertain and proportional to $ (\cos i)^{-0.5}$.

Relationship between the MCD $T_{in}$ and  $R_{in}$ parameters
and the mass and mass accretion rate 
in the Newtonian disk model with correct boundary condition is given as (see also Kubota et al.\ 1998),
\begin{equation}
M \approx 0.05 \left(\frac{T_{col}}{T_{eff}}\right)^2 (R_{in} \;{\rm km}) \;M_\odot,
\label{M}
\end{equation}
\begin{equation}
T_{in} \approx 1.1\; \left(\frac{T_{col}}{T_{eff}}\right)\left(\frac{\dot M}{10^{18}\;{\rm g\:s^{-1}}}\right)^{1/4}
\left(\frac{M}{M_\odot}\right)^{-1/2} {\rm keV}.
\label{T}
\end{equation}

The multicolor disk spectrum and the Newtonian disk with correct
boundary condition have almost identical spectral shape 
in 0.5 -- 15 keV when  equations (\ref{M}) and (\ref{T}) are considered.

Formulae to relate the new GRAD model parameters and MCD parameters
are the following:
\begin{equation}
M \approx \left(\begin{array}{c}
           0.031\\
           0.039\\
           0.045\\
           0.050\\
           0.048\\
           \end{array}\right) \left(\frac{T_{col}}{T_{eff}}\right)^2 (R_{in} \;{\rm km}) \;M_\odot
 \; {\rm for}\; i=
\left(\begin{array}{c}
     0^\circ\\
     30^\circ\\
     45^\circ\\     
     60^\circ\\
     80^\circ\\
      \end{array}\right),\label{Mgrad}
\end{equation}
\begin{equation}
T_{in} \approx \left(\begin{array}{c}
           0.75\\
           0.80\\
           0.85\\
           0.92\\
           0.96\\
           \end{array}\right)
 \left(\frac{T_{col}}{T_{eff}}\right)\left(\frac{\dot M}{10^{18}\;{\rm g\:s^{-1}}}\right)^{1/4}
\left(\frac{M}{M_\odot}\right)^{-1/2} {\rm keV}
 \; {\rm for}\; i=
\left(\begin{array}{c}
     0^\circ\\
     30^\circ\\
     45^\circ\\     
     60^\circ\\
     80^\circ\\
      \end{array}\right).\label{Tgrad}
\end{equation}
The numerical coefficients have been obtained 
by fitting the new GRAD spectra with multicolor disk model (see below) \footnote{
Equations (3) in Ebisawa, Mitsuda and Hanawa (1991) was 
wrong such that the old formula
gives 1.4 times larger mass for the same 
$R_{in}$ and $T_{in}$.}.

\begin{figure}
\begin{center}
\includegraphics[angle=-90,width=0.8\columnwidth]{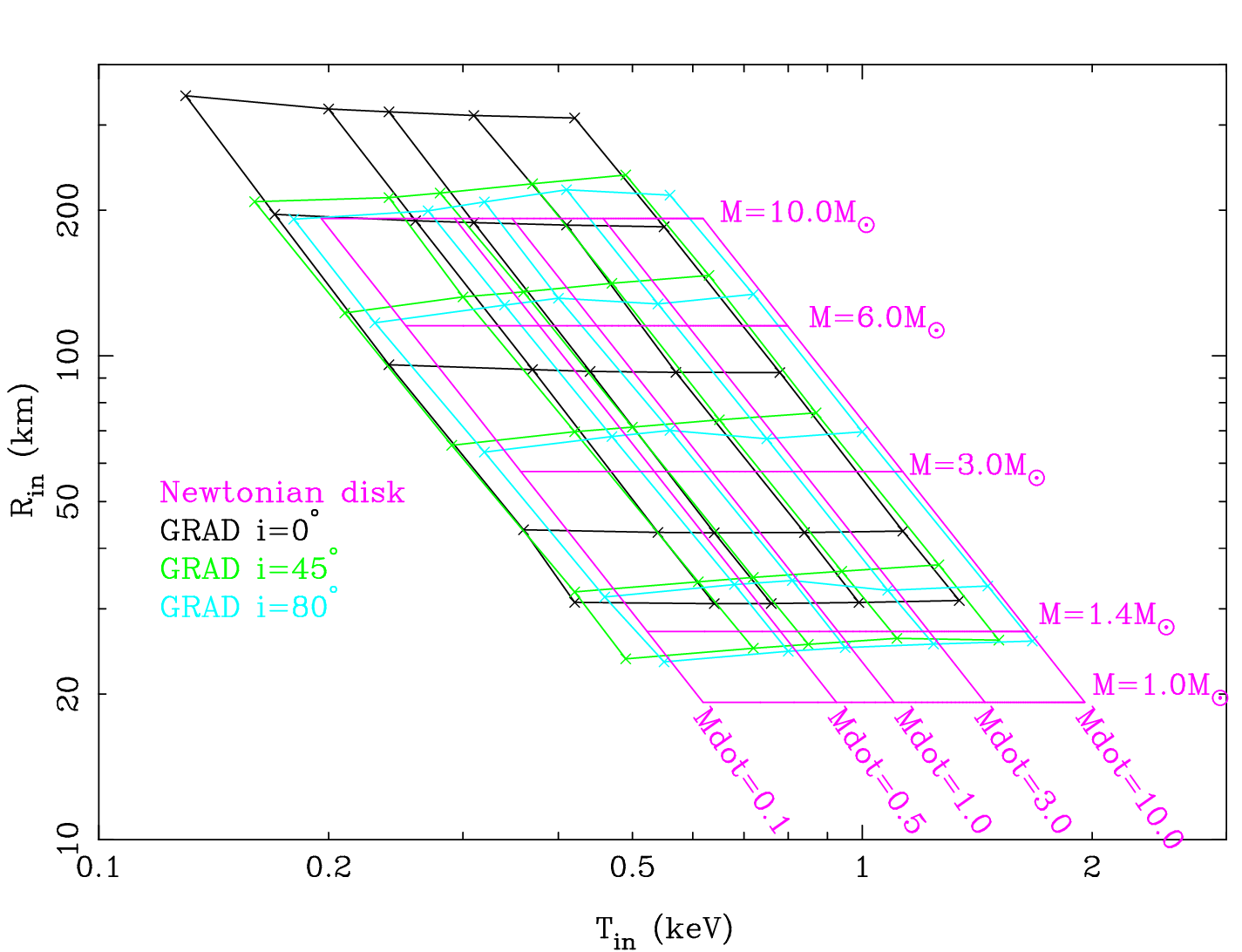}
\includegraphics[angle=-90,width=0.8\columnwidth]{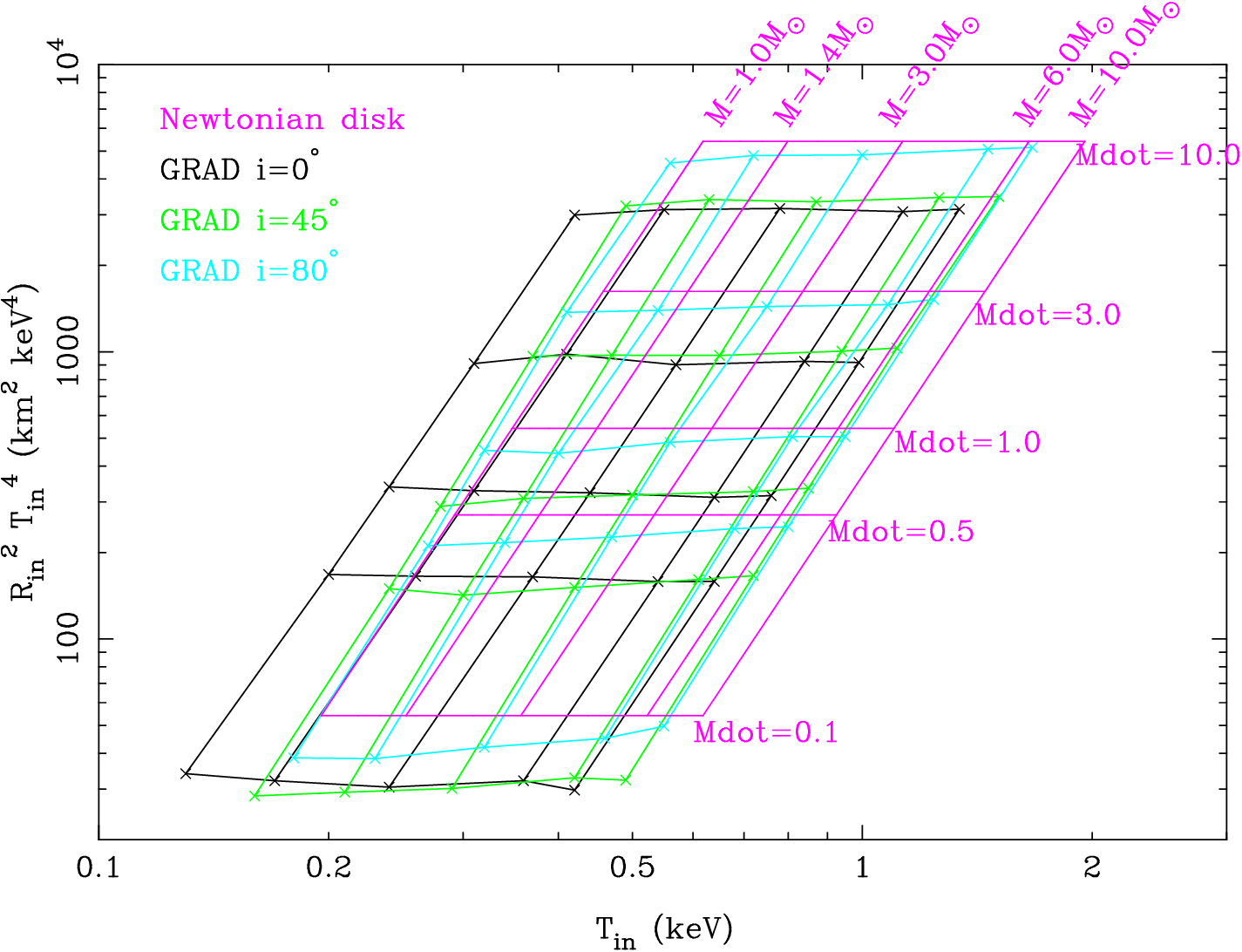}
\end{center}
\caption{Comparison of the  Newtonian and Schwarzschild disk spectra through representing
these spectra with MCD model parameters.  Upper figure shows $R_{in}$ and $T_{in}$ values to fit each
of the Newtonian or Schwarzschild disk spectra with various mass,  mass
accretion rate and inclination. Note that $R_{in}$ is proportional to mass.  In the lower figure, 
ordinate is $R_{in}^2  T_{in}^4$, which is proportional to the disk luminosity and
mass accretion rate.  Unit of the mass accretion rate is $10^{18}$ g s$^{-1}$.
Distance is assumed to be 1 kpc, and $T_{col}/T_{eff}=1$.
}\label{NewtonSchwarzschildDISKBB}
\end{figure}

In Figure \ref{NewtonSchwarzschildDISKBB}, we illustrate the relationship of the MCD model
parameters and $M$ and $\dot M$ in the Newtonian and GRAD models. These figures
are made by simulating GRAD and Newtonian model spectra with various values
of mass, mass accretion rate and inclination,  using the $Ginga$ response
(sensitive in 1 -- 30 keV), assuming the  distance  1 kpc
and exposure time  3000 sec exposure.  These simulated spectra
are fitted with the MCD model, and the best-fit $R_{in}$ and $T_{in}$ are obtained.

From equations (\ref{M}) to (\ref{Tgrad}) and
Figure \ref{NewtonSchwarzschildDISKBB}, for a given observed spectrum for which  $R_{in}$ and $T_{in}$
are determined, 
we can see the following (see also table \ref{GRO1655fit} and \ref{IC342fit}):  
\begin{itemize}

\item  When fitting the same observed spectra, we will get a smaller mass 
and higher accretion rate with the 
GRAD model than with the Newtonian disk. 

\item  Differences in the mass and mass accretion rate between the GRAD
model and the Newtonian models are the largest  for
the face-on disk, and get smaller with inclination.
\end{itemize}

These characteristics of the GRAD model are due to relativistic effects, and
may be understood intuitively as follows.  Because the conversion efficiency
of the GRAD model (0.057) is smaller than that of the Newtonian
disk (0.083), the GRAD model requires a  higher mass accretion rate
to shine with the same  luminosity. Thus we need a higher
mass accretion rate with the GRAD model when fitting the same observed spectra.
The observed disk flux is proportional to $L_{disk} \cos i$, thus the
disk luminosity induced from the observed flux is $L_{disk} \propto (\cos i)^{-1}$.
Consequently,   the mass accretion rate is proportional to 
 $(\cos i)^{-1}$ in the Newtonian case.  In the GRAD model,  this
dependence becomes  milder, because the  face-on disk
tends to  be dimmer due to  the light bending effect in the vicinity of the black hole,  while those photons removed
from the near face-on disks turn out to be the flux enhancement for the inclined disks.  In fact,
for the face-on disk, the GRAD mass accretion rate is $\sim$ 1.8 time larger
than that of the Newtonian disk, whereas it is $\sim$ 1.1 times for $i=80^\circ$. 

Because of the gravitational redshift, the GRAD model gives 
a smaller characteristic disk temperature than the Newtonian disk with the same
mass and mass accretion rate (compare equations \ref{T} and \ref{Tgrad}).
To achieve the same characteristic disk temperature of the
Newtonian model,  the $\sim$1.8  times  mass accretion rate increase  is not sufficient, 
and the mass has to get smaller  too.
Since  we measure the projected disk area $\propto M^2 \cos i$, the mass increases with inclination as $(\cos i)^{-0.5}$ 
in the Newtonian case, while
the GRAD disk temperature gets higher because of the Doppler boost.  Therefore, 
ratio of the  GRAD mass  to the Newtonian mass is the smallest for the face-on disk
($\sim$0.6) and gets larger with inclination ($\sim$0.73  for $i \gtrsim 60^\circ$).

\begin{figure}
\centerline{
\includegraphics[angle=-90,width=\columnwidth]{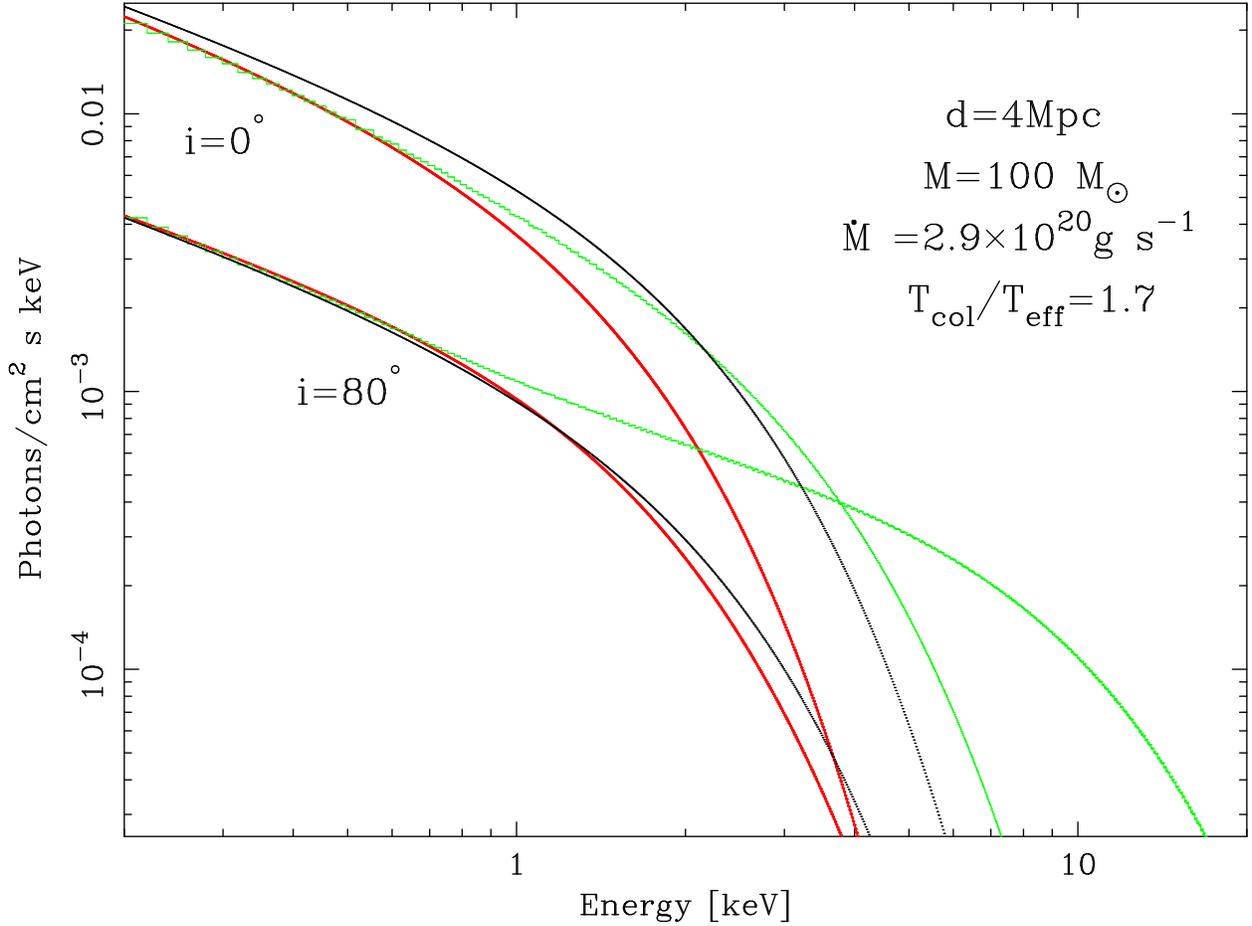} 
}
\caption{Comparison of Newtonian (black), Schwarzschild (red) 
and extreme Kerr ($a=0.998$; green) disk 
spectra for the face-on  ($i=0^\circ$) 
and a near edge-on disk ($i=80^\circ$).
 Inner disk radius is  $6 \: r_g$
for the Newtonian and Schwarzschild disks, and $1.24 \: r_g$ for the Kerr
disk.
The same mass accretion rate is assumed,
which is so chosen to give the Eddington luminosity for the Schwarzschild disk.
Other disk  parameters  are indicated  in the figure.
Note that the total disk luminosities are different for the three disk models because
the energy conversion efficiencies are  different (0.083, 0.057 and 0.366
for Newtonian, Schwarzschild and Kerr disks, respectively).
}\label{Schwarzschild_Kerr}
\end{figure}

\begin{figure}
\includegraphics[angle=-90,width=\columnwidth]{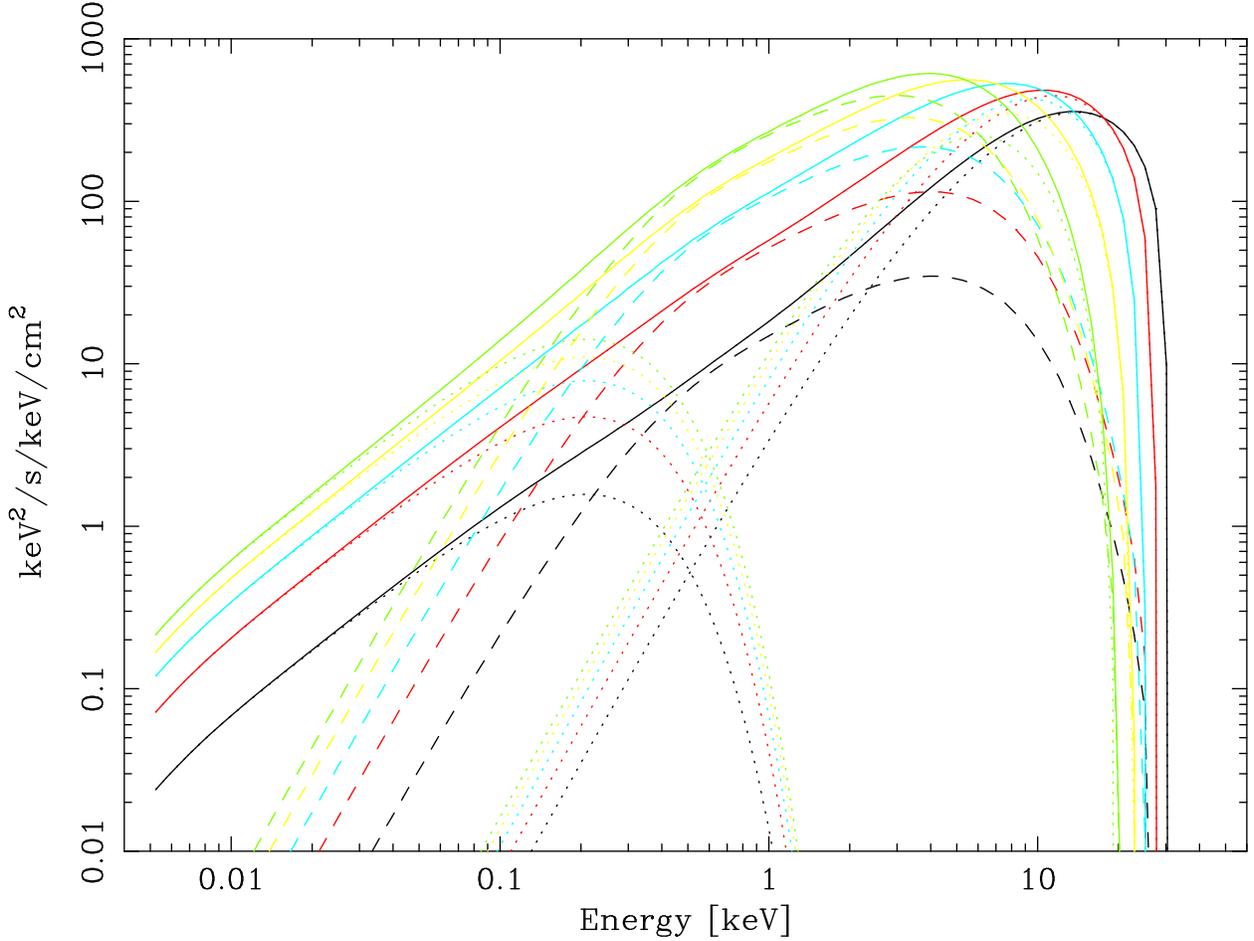}
\caption{The Kerr accretion 
disk spectra  with an extreme angular momentum (\,$a=0.998$), observed at the 
inclination angle $\mu \equiv \cos i$ = 0.9 (green; near face-on),
0.7 (yellow), 0.5 (cyan), 0.3 (red) and 0.1 (black; near edge-on).
Note the unit of the ordinate (keV$^2$ s$^{-1}$ keV$^{-1}$ cm$^{-2}$)  which facilitates
 to  see the energy release per logarithmic energy.
Solid lines indicate the total disk spectra, and contributions
from inner (\,$1.26\, r_g < r < 7 \, r_g$), 
middle (\,$7 \, r_g < r < 400 \, r_g$),  and outer parts 
(\,$400 \,r_g < r$) are plotted separately
either by dotted  line or broken line.
The distance and  mass
are assumed to be 1 kpc and  1 $M_\odot$ respectively.
The  Eddington luminosity is assumed, and $T_{col}/T_{eff}$ = 1.
}\label{kerrdisk}
\end{figure}

It is interesting that difference of the Newtonian and GRAD model parameters
are the largest  for the face-on disk and the smallest for highly inclined disks 
where relativistic effects are supposed to be most significant.
This is because the gravitational redshift and Doppler boosts
more or less cancel each other for a highly inclined  Schwarzschild disk.
This is in  contrast to the extreme Kerr disk, where near edge-on
disks show extremely hard spectra due to enormous  Doppler boosts
(see figure \ref{Schwarzschild_Kerr} and \ref{kerrdisk}).




\clearpage

\begin{deluxetable}{ccccccccccccccccccccc}
\rotate
\tabletypesize{\tiny}
\tablecaption{GRO J1655--40 fit with Newtonian,  Schwarzschild and Kerr disk models\label{GRO1655fit}\tablenotemark{a}}
\tablehead{
\colhead{}   &\multicolumn{6}{c}{Newtonian disk }&\colhead{}&\multicolumn{6}{c}{Schwarzschild disk }&\colhead{}&\multicolumn{6}{c}{Kerr disk  ($a=0.998$)}\\
             \cline{2-7} \cline{9-14} \cline{16-21}\\ 
\colhead{$i$}&\colhead{$M/M_\odot$}&\colhead{$\dot M$\tablenotemark{b}}&\colhead{$\dot M/\dot M_C$\tablenotemark{c}}&\colhead{$N_{pow}$\tablenotemark{d}}&\colhead{$N_H$\tablenotemark{e}}&\colhead{$\chi2$/dof}&\colhead{}
             &\colhead{$M/M_\odot$}&\colhead{$\dot M$\tablenotemark{b}}&\colhead{$\dot M/\dot M_C$\tablenotemark{c}}&\colhead{$N_{pow}$\tablenotemark{d}}&\colhead{$N_H$\tablenotemark{e}}&\colhead{$\chi2$/dof}&\colhead{}
             &\colhead{$M/M_\odot$}&\colhead{$\dot M$\tablenotemark{b}}&\colhead{$\dot M/\dot M_C$\tablenotemark{c}}&\colhead{$N_{pow}$\tablenotemark{d}}&\colhead{$N_H$\tablenotemark{e}}&\colhead{$\chi2$/dof}}
\startdata
0$^\circ$   &1.3&0.49 &0.18&7.2 &0.97&1.3   & & 0.83  & 0.87& 0.37   & 6.9    &0.97 &1.35&&2.6  & 0.58 &0.48 &6.2&0.98 &1.40\\
30$^\circ$  &1.4&0.56 &0.19&7.2 &0.97&1.3   & & 0.96  & 0.99& 0.36   & 6.0    &0.95 &1.25&&3.0  & 0.33 &0.24 &4.0&0.94 &1.20\\
45$^\circ$  &1.6&0.69 &0.22&7.2 &0.97&1.3   & & 1.2   & 1.2 & 0.34   & 5.2    &0.93 &1.17&&5.2  & 0.36 &0.15 &3.9&0.94 &1.30\\
70$^\circ$  &2.3&1.42 &0.31&7.2 &0.97&1.3   & & 1.8   & 2.1 & 0.40   & 4.7    &0.92 &1.08&&15.9 & 0.35 &0.048&8.9&0.99 &1.45\\
80$^\circ$  &3.2&2.80 &0.43&7.2 &0.97&1.3   & & 2.4   & 3.1 & 0.45   & 6.3    &0.93 &1.14&&27.0 & 0.39 &0.032&10.4&0.99 &1.62\\
\enddata
\tablenotetext{a}{Distance is assumed to be 3.2 kpc, and $T_{col}/T_{eff}$=1.7. Slope of the power-law component is fixed to 2.5. One percent systematic error is included for each spectral bin.  Degree of freedom (dof) is 156.}
\tablenotetext{b}{Unit is $10^{18}$ g s$^{-1}$.}
\tablenotetext{c}{The critical mass accretion rate $\dot M_C$ is so defined that $\dot M = \dot M_C$ gives the Eddington luminosity.
                  See Appendix \protect\ref{efficiency_appendix}.}
\tablenotetext{d}{Power-law component normalization in photons s$^{-1}$ cm$^{-2}$ keV$^{-1}$ at 1 keV.}
\tablenotetext{e}{In $10^{22}$ cm$^{-2}$.}
\end{deluxetable}

\begin{deluxetable}{cccccccccccccccccc}
\rotate
\tabletypesize{\footnotesize}
\tablecaption{IC342 Source 1 fit with Newtonian,  Schwarzschild and Kerr disk models\label{IC342fit}\tablenotemark{a}}
\tablehead{
\colhead{}   &\multicolumn{5}{c}{Newtonian disk }&\colhead{}&\multicolumn{5}{c}{Schwarzschild disk }&\colhead{}&\multicolumn{5}{c}{Kerr disk  ($a=0.998$)}\\
             \cline{2-6} \cline{8-12} \cline{14-18}\\ 
\colhead{$i$}&\colhead{$M/M_\odot$}&\colhead{$\dot M$\tablenotemark{b}}&\colhead{$\dot M/\dot M_C$\tablenotemark{c}}&\colhead{$N_H$\tablenotemark{d}}&\colhead{$\chi2$/dof}&\colhead{}
             &\colhead{$M/M_\odot$}&\colhead{$\dot M$\tablenotemark{b}}&\colhead{$\dot M/\dot M_C$\tablenotemark{c}}&\colhead{$N_H$\tablenotemark{d}}&\colhead{$\chi2$/dof}&\colhead{}
             &\colhead{$M/M_\odot$}&\colhead{$\dot M$\tablenotemark{b}}&\colhead{$\dot M/\dot M_C$\tablenotemark{c}}&\colhead{$N_H$\tablenotemark{d}}&\colhead{$\chi2$/dof}}
\startdata
0$^\circ$   &14.6&170 &5.8 &0.42&0.91&   &  8.9   & 304 & 11.8   & 0.43   & 0.90 &&27.3  & 195 &15.5 &0.50 &0.91\\
30$^\circ$  &15.7&197 &6.3 &0.42&0.91&   & 10.1   & 335 & 11.4   & 0.44   & 0.90 &&29.5  & 106 &7.8  &0.51 &0.90\\
45$^\circ$  &17.4&241 &6.9 &0.41&0.91&   & 11.9   & 387 & 11.2   & 0.44   &0.90  &&51.3  & 113 &4.8  &0.51 &0.90\\
60$^\circ$  &20.7&341 &8.2 &0.42&0.91&   & 15.2   & 502 & 11.4   & 0.44   &0.89  &&107 & 120 &2.4  &0.46 &0.91 \\
80$^\circ$  &35.1&981 &14.0&0.42&0.91&   & 25.9   & 1069& 14.2   & 0.38   &0.91  &&332   & 153 &1.0  &0.30 &1.00\\
\enddata
\tablenotetext{a}{Distance is assumed to be 4 Mpc, and $T_{col}/T_{eff}$=1.7. Degree of freedom (dof) is 52.}
\tablenotetext{b}{Unit is $10^{18}$ g s$^{-1}$.}
\tablenotetext{c}{The critical mass accretion rate $\dot M_C$ is so defined that $\dot M = \dot M_C$ gives the Eddington luminosity.
                 See Appendix \protect\ref{efficiency_appendix}.}
\tablenotetext{d}{In $10^{22}$ cm$^{-2}$.}
\end{deluxetable}

\begin{deluxetable}{ccccccccc}
\rotate
\tablecaption{IC342 Source 1 fit with Slim disk model\label{IC342fitslim}\tablenotemark{a}}
\tablehead{\colhead{$i$}&\colhead{$M/M_\odot$}&\colhead{$\dot M$\tablenotemark{b}}&\colhead{$\dot M/\dot M_C$\tablenotemark{c}}&\colhead{$L_{disk}/L_{Edd}$}&\colhead{$N_H$\tablenotemark{d}}&\colhead{$\chi2$/dof}}
\startdata
0$^\circ$   &$19.8\pm1.7$&$1000\pm^{1300}_{800}$ & $20\pm^{23}_{16}$&5.9 &$0.90\pm^{0.08}_{0.06}$&1.40  \\ 
\enddata
\tablenotetext{a}{Distance is assumed to be 4 Mpc, and $T_{col}/T_{eff}$=1.7. The
viscous parameter $\alpha=0.01$. Degree of freedom (dof) is 52.  Errors correspond to 90 \% confidence level.}
\tablenotetext{b}{Unit is $10^{18}$ g s$^{-1}$.}
\tablenotetext{c}{The critical mass accretion rate $\dot M_C$ is so defined that $\dot M = \dot M_C$ gives the Eddington luminosity in the case of the standard disk.
                 See Appendix \protect\ref{efficiency_appendix}.}
\tablenotetext{d}{In $10^{22}$ cm$^{-2}$.}
\end{deluxetable}
\clearpage

\clearpage

\clearpage

\clearpage

\clearpage

\end{document}